\documentclass[a4paper,USenglish,cleveref, autoref]{lipics-v2019}

\title{Non-deterministic weighted automata \protect\\ evaluated over Markov chains\footnote{This paper has been published in Journal of Computer and System Sciences: \url{https://doi.org/10.1016/j.jcss.2019.10.001}. }}
\titlerunning{Non-deterministic weighted automata evaluated over Markov chains}

\author{Jakub Michaliszyn}{University of Wroc\l{}aw}{jmi@cs.uni.wroc.pl}{https://orcid.org/0000-0002-5053-0347}{} 
\author{Jan Otop}{University of Wroc\l{}aw}{jotop@cs.uni.wroc.pl}{https://orcid.org/0000-0002-8804-8011}{}

\authorrunning{Jakub Michaliszyn and Jan Otop}

\Copyright{Jakub Michaliszyn and Jan Otop}

\ccsdesc[500]{Theory of computation~Automata over infinite objects}
\ccsdesc[500]{Theory of computation~Quantitative automata} 

\keywords{quantitative verification, weighted automata,expected value}

\usepackage[
    type={CC},
    modifier={by-nc-nd},
    version={4.0},
]{doclicense}

\nolinenumbers 

\hideLIPIcs

\funding{The National Science Centre (NCN), Poland under grant 2017/27/B/ST6/00299.}

\clearpage{}\usepackage{amssymb,amsthm,amsmath}
\usepackage{tikz}
\usepackage{environ}
\usepackage{lineno}

\usetikzlibrary{arrows,automata,shapes,positioning}

\newtheorem{fact}[theorem]{Fact}
\newtheorem{nremark}[theorem]{Remark}
\newcommand{\set}[1]{\{#1\}}
\newcommand{\std}{\mathrm{std}}

\newcommand{\aut}{\mathcal{A}}
\newcommand{\Q}{\mathbb{Q}}
\newcommand{\Z}{\mathbb{Z}}
\newcommand{\N}{\mathbb{N}}
\newcommand{\R}{\mathbb{R}}
\newcommand{\flimavg}{\textsc{LimAvg}}
\newcommand{\fsum}{\textsc{Sum}}
\newcommand{\fmax}{\textsc{Max}}
\newcommand{\fmin}{\textsc{Min}}
\newcommand{\fsup}{\textsc{Sup}}
\newcommand{\finf}{\textsc{Inf}}
\newcommand{\favgFin}{\textsc{Avg}}
\newcommand{\favg}[1]{\textsc{Avg}(#1)}
\newcommand{\rev}[1]{\frac{1}{#1}}

\newcommand{\Paragraph}[1]{\noindent\textbf{#1.}}

\newcommand{\PSPACE}{\textsc{PSpace}{}}

\newcommand{\prob}{\mathbb{P}}
\newcommand{\expected}{\mathbb{E}}
\newcommand{\valueL}[1]{\mathcal{L}_{{#1}}}

\newcommand{\cost}{{C}}
\newcommand{\markov}{\mathcal{M}}

\newcommand{\distrib}{\mathbb{D}}
\newcommand{\const}{\lambda}
\newcommand{\tuple}[1]{\langle #1 \rangle}

\newcommand{\sharpP}{\textsc{\#P}}

\newcommand{\autInf}{\aut_{\infty}}

\begin{document}
\maketitle

\begin{abstract}
We present the first study of non-deterministic weighted automata under probabilistic semantics.
In this semantics words are random events, generated by a Markov chain, and functions computed by weighted automata are random variables. 
We consider the probabilistic questions of computing the expected value and the cumulative distribution for such random variables.

The exact answers to the probabilistic questions for non-deterministic automata can be irrational and are uncomputable in general. 
To overcome this limitation, we propose approximation algorithms for the probabilistic questions, which work in exponential time in the size of the automaton and polynomial time in the size of the Markov chain and the given precision.
We apply this result to show that non-deterministic automata can be effectively determinised with respect to the standard deviation metric.

 \end{abstract}

\newcommand{\introPara}[1]{\noindent\emph{#1}.}
\section{Introduction}

Weighted automata are (non-deterministic) finite automata in which transitions carry weights~\cite{Droste:2009:HWA:1667106}. 
We study here weighted automata (on finite and infinite words) whose semantics is given by \emph{value functions} (such as the sum or the average)~\cite{quantitativelanguages}. 
In such weighted automata transitions are labeled with rational numbers and hence every run yields a sequence of rationals, which the value function aggregates into a single 
(real) number. 
This number is the value of the run, and 
the value of a word is the infimum over the values of all accepting runs on that word. 

The value function approach has been introduced to express quantitative system properties (performance, energy consumption, etc.) and it serves as a foundation for \emph{quantitative verification}~\cite{quantitativelanguages,henzingerotop17}.
Basic decision questions for weighted automata are quantitative counterparts of the emptiness and universality questions obtained by imposing a threshold on the values of words.

\introPara{Probabilistic semantics}
The emptiness and the universality problems correspond to the best-case and the worst-case analysis. For the average-case analysis, weighted automata are considered under probabilistic semantics, in which words are random events 
generated by a Markov chain~\cite{DBLP:conf/icalp/ChatterjeeDH09,lics16}. 
In such a setting, functions from words to reals computed by deterministic weighted automata are measurable and hence can be considered as random variables. 
The fundamental probabilistic questions are to compute \emph{the expected value} and \emph{the cumulative distribution} for a given automaton and a Markov chain.

\smallskip
\introPara{The deterministic case} 
Weighted automata under probabilistic semantics have been studied only in the deterministic case. 
A close relationship has been established between weighted automata under probabilistic semantics and weighted Markov chains~\cite{DBLP:conf/icalp/ChatterjeeDH09}.
For a weighted automaton $\aut$ and a Markov chain $\markov$ representing the distribution over words, the probabilistic problems for $\aut$ and $\markov$ coincide with the probabilistic problem of the weighted Markov chain $\aut \times \markov$.
Weighted Markov chains have been intensively studied with single and multiple quantitative objectives~\cite{BaierBook,DBLP:conf/concur/ChatterjeeRR12,filar,DBLP:conf/cav/RandourRS15}.
 The above reduction does not extend to non-deterministic weighted automata \cite[Example~30]{lics16}. 

\smallskip
\introPara{Significance of nondeterminism} 
Non-deterministic weighted automata are provably more expressive than their deterministic counterpart~\cite{quantitativelanguages}. 
Many important system properties 
can be expressed with weighted automata only in the nondeterministic setting. This includes minimal response time, minimal number of errors and the edit distance problem~\cite{henzingerotop17}, which serves as the foundation for the \emph{specification repair} framework from~\cite{DBLP:conf/lics/BenediktPR11}.

Non-determinism can also arise as a result of abstraction. The exact systems are often too large and complex to operate on and hence they are approximated with 
smaller non-deterministic models~\cite{clarke2016handbook}. The abstraction is especially important for multi-threaded programs, where the explicit model grows exponentially with the number of threads~\cite{DBLP:conf/popl/GuptaPR11}. 

\paragraph*{Our contributions}
We study non-deterministic weighted automata under probabilistic semantics. 
We work with weighted automata as defined in~\cite{quantitativelanguages}, where a value function $f$ is used to aggregate weights along a run, and
the value of the word is the infimum over values of all runs. (The infimum can be changed to supremum as both definitions are dual).
We primarily focus on the two most interesting value functions: the sum of weights over finite runs, and
the limit average over infinite runs.
The main results presented in this paper are as follows.
\begin{itemize}
\item We show that the answers to the probabilistic questions for weighted automata with the sum and limit-average value functions can be irrational and even transcendental (Theorem~\ref{th:irrational}) and cannot be computed by any effective representation (Theorem~\ref{th:limavg-undecidable}). 

\item We establish approximation algorithms for the probabilistic questions for weighted automata with the sum and limit-average value functions.
The approximation is \sharpP-complete for (total) weighted automata with the sum value function (Theorem~\ref{th:approximation-sum}), and it 
is $\PSPACE$-hard and solvable in exponential time for weighted automata with the limit-average value function (Theorem~\ref{th:approximation-limavg}).

\item We show that weighted automata with the limit-average value function can be approximately determinised (Theorem~\ref{th:approximateDeterminisation}). 
Given an automaton $\aut$
and $\epsilon >0$, we show how to compute a deterministic automaton $\aut_D$ such that the expected difference between the values returned by both automata is at most $\epsilon$.
\end{itemize}

\paragraph*{Applications}
We briefly discuss applications of our contributions in quantitative verification.
\begin{itemize}

\item The expected-value question corresponds to the average-case analysis in quantitative verification~\cite{DBLP:conf/icalp/ChatterjeeDH09,lics16}.  
Using results from this paper, we can perform the average-case analysis with respect to quantitative specifications given by non-deterministic weighted automata. 

\item Some quantitative-model-checking frameworks~\cite{quantitativelanguages} are based on the universality problem for non-deterministic automata, which asks whether all words have the value below a given threshold. 
Unfortunately, the universality problem is undecidable for weighted automata with the sum or the limit average values functions. 
The distribution question can be considered as a computationally-attractive variant of universality, i.e., we ask whether almost all words have value below some given threshold. 
We show that if the threshold can be approximated, the distribution question can be computed effectively.

\item Weighted automata have been used to formally study online algorithms~\cite{aminof2010reasoning}. Online algorithms have been modeled by deterministic weighted automata, which  make choices based solely on the past, while
offline algorithms have been modeled by non-deterministic weighted automata. 
Relating deterministic and non-deterministic models allowed for formal verification of the worst-case competitiveness ratio of online algorithms.
Using the result from our paper, we can extend the analysis from~\cite{aminof2010reasoning} to the average-case competitiveness.
\end{itemize}

\paragraph*{Related work} The problem considered in this paper is related to the following areas from the literature.

\noindent\emph{Probabilistic verification of qualitative properties}.
Probabilistic verification asks for the probability of the set of traces satisfying a given property. 
For non-weighted automata, it has been extensively studied~\cite{DBLP:conf/focs/Vardi85,DBLP:journals/jacm/CourcoubetisY95,
BaierBook} and implemented~\cite{DBLP:journals/entcs/KwiatkowskaNP06,DBLP:conf/tacas/HintonKNP06}. 
The prevalent approach in this area is to work with deterministic automata, and apply determinisation as needed. 
To obtain better complexity bounds, the probabilistic verification problem has been directly studied for unambiguous B\"uchi automata in~\cite{DBLP:conf/cav/BaierK0K0W16};
the authors explain there the potential pitfalls in the probabilistic analysis of non-deterministic automata. 

\smallskip
\noindent\emph{Weighted automata under probabilistic semantics}.
Probabilistic verification of weighted automata and their extensions has been studied in~\cite{lics16}.
All automata considered there are deterministic.

\smallskip
\noindent\emph{Markov Decision Processes (MDPs)}. MDPs are a classical extension of Markov chains, which models control in a stochastic environment~\cite{BaierBook,filar}.
In MDPs, probabilistic and non-deterministic transitions are interleaved;
this can be explained as a game between two players: Controller and Environment. 
Given a game objective (e.g. state reachability), the goal of Controller is to maximize the probability of the objective by selecting non-deterministic transitions.
Environment is selecting probabilistic transitions at random w.r.t. a probability distribution described in the current state of the MDP.
Intuitively, the non-determinism in MDPs is resolved based on the past, i.e., each time Controller selects a non-deterministic transition, its choice is based on previously picked transitions.
Our setting can be also explained in such a game-theoretic framework: first, Environment generates a complete word, and only then non-deterministic choices are resolved by Controller, who generates a run of a given weighted automaton. 
That non-determinism in a run at some position $i$ may depend on letters in the input word on positions past $i$ (i.e., future events).

Partially Observable Markov Decision Process (POMDPs)~\cite{ASTROM1965174} are an extension of MDPs, which models weaker non-determinism.
In this setting, the state space is partitioned into \emph{observations} and the non-deterministic choices have to be the same for sequences consisting of the same observations (but possibly different states).
Intuitively, Controller can make choices based only on the sequence of observations it has seen so far.
While in POMDPs Controller is restricted, in our setting Controller is stronger than in the MDPs case.

\smallskip
\noindent\emph{Non-deterministic probabilistic automata}.
The combination of nondeterminism with stochasticity has been recently studied in the framework of probabilistic automata~\cite{nondet-prob}.
There have been defined non-deterministic probabilistic automata (NPA) and there has been proposed two possible semantics for NPA. 
It has been shown that the equivalence problem for NPA is undecidable (under either of the considered two semantics).
Related problems, such as the threshold problem, are undecidable already for (deterministic) probabilistic automata~\cite{bertoni1977some}. 
While NPA work only over finite words, the interaction between probabilistic and non-deterministic transitions is more general than in our framework.
In particular, non-determinism in NPA can influence the probability distribution, which is not possible in our framework.

\smallskip
\noindent\emph{Approximate determinisation}.
As weighted automata are not determinisable, Boker and Henzinger~\cite{BokerH12} studied \emph{approximate} determinisation defined as follows.
The distance $d_{\sup}$ between weighted automata $\aut_1, \aut_2$ is defined as  $d_{\sup}(\aut_1, \aut_2) = \sup_{w} | \aut_1(w) - \aut_2(w)|$. 
A nondeterministic weighted automaton $\aut$ can be \emph{approximately} determinised if for every $\epsilon >0$
there exists a deterministic automaton $\aut_D$ such that $d_{\sup}(\aut, \aut_D) \leq \epsilon$. 
Unfortunately, weighted automata with the limit average value function cannot be approximately determinised~\cite{BokerH12}.
In this work we show that the approximate determinisation is possible for the standard deviation metric $d_{\std}$
defined as $d_{\std}(\aut_1, \aut_2) = \expected(|\aut_1(w) - \aut_2(w)|)$.
\medskip

This paper is an extended and corrected version of~\cite{concur2018}. It contains full proofs, an extended discussion and a stronger version of  Theorem~\ref{th:irrational}.
We have showed in~\cite{concur2018} that the expected values and the distribution values may be irrational. 
In this paper we show that these values can be even transcendental (Theorem~\ref{th:irrational}).

We have corrected two claims from~\cite{concur2018}.
First, we have corrected statements of Theorems~\ref{th:irrational} and~\ref{th:limavg-undecidable}. For $\flimavg$-automata and the distribution question $\distrib_{\markov,\aut}(\lambda)$, 
the values, which can be irrational and uncomputable are not the values of the probability $\distrib_{\markov,\aut}(\lambda) = \prob_{\markov}(\set{w \mid \valueL{\aut} \leq \lambda})$, but the values of the thershod $\lambda$ that correspond to mass points, i.e., 
values $\lambda$ such that $\prob_{\markov}(\set{w \mid \valueL{\aut} = \lambda}) > 0$.
We have also removed from Theorem~\ref{th:all-pspace-hard} PSPACE-hardness claim for the distribution question for (non-total) $\fsum$-automata. 
We show that the (exact) distribution question for all $\fsum$-automata is \sharpP-complete.
 
\section{Preliminaries}

Given a finite alphabet $\Sigma$ of letters, a \emph{word} $w$ is a finite or infinite sequence 
of letters.
We denote the set of all finite words over $\Sigma$ by $\Sigma^*$, and the set of all infinite words over $\Sigma$ by $\Sigma^\omega$.
For a word $w$, we define $w[i]$ as the $i$-th letter of $w$, and we define $w[i,j]$ as the subword $w[i] w[i+1] \ldots w[j]$ of $w$. 
We use the same notation for other sequences defined later on.
By $|w|$ we denote the length of $w$. 

A \emph{(non-deterministic) finite automaton} (NFA) is a tuple $(\Sigma, Q, Q_0, F, \delta)$
consisting of
	an input alphabet $\Sigma$ , 
	a finite set of states $Q$, 
	a set of initial states $Q_0 \subseteq Q$,  
	a set of final states $F$, and
    a finite transition relation 	$\delta \subseteq Q \times \Sigma \times Q$.

We define $\delta(q,a) = \set{q' \in \Q \mid \delta(q,a,q')}$
and $\delta(S,a) = \bigcup_{q \in S} \delta(q,a)$.
We extend this to words $\widehat{\delta} \colon 2^Q \times \Sigma^* \to 2^Q$ in the following way:
$\widehat{\delta}(S,\epsilon) = S$ (where $\epsilon$ is the empty word) and 
 $\widehat{\delta}(S,aw) = \widehat{\delta}(\delta(S,a),w)$, i.e., $\widehat{\delta}(S,w)$ is the set of states reachable from $S$ via $\delta$ over the word $w$. 

\Paragraph{Weighted automata}
A \emph{weighted automaton} is a finite automaton whose transitions are labeled by rational numbers called \emph{weights}. 
Formally, a weighted automaton is a tuple 
$(\Sigma, Q, Q_0, F, \delta, \cost)$, where the first five elements are as in the finite automata, and $\cost \colon \delta \to \mathbb{Q}$ is a function that defines \emph{weights} of transitions. 
An example of a weighted automaton is depicted in Figure~\ref{fig:aut}.

The size of a weighted automaton $\aut$, denoted by $|\aut|$,
 is $|Q| + |\delta| + \sum_{q, q', a} \mathrm{len}(C(q, a, q'))$, where $\mathrm{len}$ is the sum of the lengths of the binary representations of the numerator and the denominator of a given rational number.

A \emph{run} $\pi$  of an automaton $\aut$ on a word $w$ is a sequence of states $\pi[0] \pi[1] \dots$ such that $\pi[0]$ is an initial state and for each $i$ we have $(\pi[i-1],w[i],\pi[i]) \in \delta$.
A finite run $\pi$ of length $k$ is \emph{accepting} if and only if the last state $\pi[k]$ belongs to the set of accepting states $F$.
As in~\cite{quantitativelanguages}, we do not consider $\omega$-accepting conditions and assume that all infinite runs are accepting.
Every run $\pi$ of an automaton $\aut$ on a (finite or infinite) word $w$ defines a sequence of weights 
of successive transitions of $\aut$ as follows.
Let $(\cost(\pi))[i]$ be the weight of the $i$-th transition,
i.e., $\cost(\pi[i-1], w[i], \pi[i])$. 
Then, $\cost(\pi)=(\cost(\pi)[i])_{1\leq i \leq |w|}$.
A \emph{value functions} $f$ is a function that 
assigns real numbers to sequences of rational numbers.
The value $f(\pi)$ of the run $\pi$ is defined as $f(\cost(\pi))$.

The value of a (non-empty) word $w$ assigned by the automaton $\aut$, denoted by $\valueL{\aut}(w)$,
is the infimum of the set of values of all accepting runs on $w$. The value of a word that has no (accepting) runs is infinite.
To indicate a particular value function $f$ that defines the semantics,
we will call a weighted automaton $\aut$ an $f$-automaton.

\Paragraph{Value functions}
We consider the following value functions. For finite runs, functions $\fmin$ and $\fmax$ are defined in the usual manner, and the function $\fsum$ is defined as 
\[\fsum(\pi) = \sum\nolimits_{i=1}^{|C(\pi)|} (\cost(\pi))[i]\]

For infinite runs we consider the supremum $\fsup$ and infimum $\finf$ functions (defined like $\fmax$ and $\fmin$ but on infinite runs) and the limit average function $\flimavg$ defined as 
\[\flimavg(\pi) = \limsup\limits_{k \rightarrow \infty} \favg{\pi[0, k]} \]
where for finite runs $\pi$ we have \(\favg{\pi}=\frac{\fsum(\pi)}{|C(\pi)|}\).

\subsection{Probabilistic semantics}

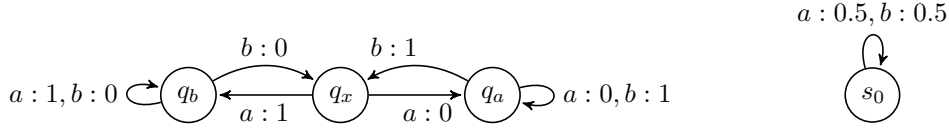
\begin{figure}
\centering
{
\centering
\begin{tikzpicture}[->,>=stealth',shorten >=1pt,auto,node distance=2.0cm,semithick]
  \tikzstyle{every state}=[fill=white,draw=black,text=black,minimum size=0.4cm]
  \tikzset{
   datanode/.style = {
    draw,
    circle,
    text width=0.30cm,
    inner ysep = +0.4em},
    labelnode/.style = {
    draw, 
    rounded corners,
    align=center,
    fill=white}
  }

\foreach \lab\lX/\lY/\name/\kind/\cost/\time/\attr in 
         {B/-4/1/$q_b$/square/0/10/5,
          X/-2/1/$q_x$/walk/0/100/10,
          A/0/1/$q_a$/bus/10/15/-2,
          M/5/1/$s_0$/m/1/1/1}
{
  
  \node[datanode] (\lab) at (\lX,\lY)  
  {
   \name
  };
  
}

\draw  (X) edge node[below] {$\,\,\,\,\,a:0$} (A);
\draw  (X) edge node {$a:1$} (B);
\draw  (A) edge[bend right] node[above] {$b:1\ \ \ \ \ $}  (X);
\draw  (B) edge[bend left] node {$b:0$} (X);
\draw  (A) edge[loop right] node {$a:0, b:1$} (A);
\draw  (B) edge[loop left] node {$a:1, b:0$} (B);

\draw  (M) edge[loop above] node {$a:0.5, b:0.5$} (B);

\end{tikzpicture}
} \caption{The automaton $\aut=\set{\set{a,b}, \set{q_x, q_a, q_b}, \set{q_a, q_b}, \emptyset, \delta, \cost}$, where $\delta=\set{(q_a, a, q_a),(q_a,b, q_a), (q_a,b,q_x), (q_x, a, q_a), (q_x, a, q_b), (q_b, a, q_x), (q_b, a, q_b), (q_b,b,q_b)}$ and $C$ such that $C(q_a, b, q_a)=C(q_b,  a, q_b)=C(q_a, b, q_x)=C(q_x,a,q_b)=1$ and for all other inputs the value of $C$ is $0$ (left) and the Markov chain $\markov=\set{\set{a,b},\set{s_0},\set{s_0},E}$ where $E$ always returns $0.5$ (right).}
\label{fig:aut}
\end{figure}

A (finite-state discrete-time) \emph{Markov chain} is a tuple $\tuple{\Sigma,S,s_0,E}$, 
where $\Sigma$ is the alphabet of letters, 
$S$ is a finite set of states, $s_0$ is an initial state, 
$E \colon S \times \Sigma \times S \mapsto [0,1]$ is an edge probability function, which  
for every $s \in S$ satisfies that $\sum_{a \in \Sigma, s' \in S} E(s,a,s') = 1$. 
An example of a single-state Markov chain is depicted in Figure~\ref{fig:aut}.

In this paper, Markov chains serve as a mathematical model as well as the input to algorithms.
Whenever a Markov chain is the input to a problem or an algorithm, we assume that all edge probabilities are rational and the size of a Markov chain $\markov$ is defined as
$|\markov|=|S|+|E|+\sum_{q, q', a}\mathrm{len}(E(q, a, q'))$.

The probability of a finite word $u$ w.r.t.\ a Markov chain $\markov$, denoted by $\prob_{\markov}(u)$,  is the sum of probabilities of paths from $s_0$ labeled by $u$, 
where the probability of a path is the product of probabilities of its edges.
For sets $u\cdot \Sigma^\omega = \{ uw \mid w \in \Sigma^{\omega} \}$, called \emph{cylinders},
we have $\prob_{\markov}(u\cdot \Sigma^\omega)=\prob_{\markov}(u)$, and then the 
probability measure over infinite words defined by $\markov$ is the unique
extension of the above measure to the $\sigma$-algebra generated by cylinders (by Carath\'{e}odory's extension theorem~\cite{feller}) 
We will denote the unique probability measure defined by $\markov$ as $\prob_{\markov}$.
For example, for the Markov chain $\markov$ presented in Figure~\ref{fig:aut}, we have that $\prob_{\markov}(ab) = 
\frac{1}{4}$, and so $\prob_{\markov}(\set{w \in \set{a,b}^\omega \mid w[0,1]=ab})=\frac{1}{4}$, whereas $\prob_{\markov}(X)=0$ for any countable set of infinite words $X$.

A function $f \colon \Sigma^{\omega} \to \R$ the is measureable w.r.t.  $\prob_{\markov}$ is called a \emph{random variable} (w.r.t. $\prob_{\markov}$). 
A random variable $g$ is \emph{discrete}, if there exists a countable set $Y \subset \R$ such  that $g$ returns a value for $Y$ with probability $1$ ($\prob_{\markov}(\set{w \mid g(w) \in Y}) = 1$).
For the discrete random variable $g$, we define the \emph{expected value} $\expected_{\markov}(g)$ (w.r.t. the measure $\prob_{\markov}$) as 
\[
\expected_{\markov}(g) = \sum_{y \in Y} y \cdot \prob_{\markov}(\set{w \mid g(w) = y}).
\]
Every non-negative random variable $h \colon \Sigma^{\omega} \to \R^+$ is a point-wise limit of some sequence of monotonically increasing discrete random variables $g_1, g_2, \ldots$ and the expected value  $\expected_{\markov}(h)$ is the limit 
of expected values $\expected_{\markov}(g_i)$~\cite{feller}. Finally, every random variable $f$ can be presented as the difference $h_1 - h_2$ of non-negative random variables $h_1, h_2$ and we have
$\expected_{\markov}(f) = \expected_{\markov}(h_1) - \expected_{\markov}(h_2)$~\cite{feller}.

A \emph{terminating} Markov chain $\markov^T$ is 
a tuple $\tuple{\Sigma,S,s_0,E, T}$, 
where $\Sigma$, $S$ and $s_0$ are as usual,
$E \colon S \times (\Sigma\cup \set{\epsilon}) \times S \mapsto [0,1]$ is the edge probability function, such that if $E(s, a, t)$, then $a=\epsilon$ if and only if $t\in T$, and
for every $s \in S$ we have $\sum_{a \in \Sigma\cup \set{\epsilon}, s' \in S} E(s,a,s') = 1$, 
and 
$T$ is a set of terminating states such that the probability of reaching a terminating state from any state $s$ is positive.
Notice that the only $\epsilon$-transitions in a terminating Markov chain are those that lead to a terminating state.

The probability of a finite word $u$ w.r.t. $\markov^T$, denoted $\prob_{\markov^T}(u)$,  is the sum of probabilities of paths from $s_0$ labeled by $u$ such that the only terminating state on this path is the last one. 
Notice that $\prob_{\markov^T}$ is a probability distribution on finite words whereas $\prob_{\markov}$ is not (because the sum of probabilities may exceed 1).

A function $f \colon \Sigma^{*} \to \R$ is called a \emph{random variable} (w.r.t. $\prob_{\markov^T}$).
Since words generated by $\markov^T$ are finite, the co-domain of $f$ is countable and hence $f$ is discrete. The expected value of $f$ w.r.t. $\markov^T$ is defined in the same way as for non-terminating Markov chains. 

\Paragraph{Automata as random variables}
An infinite-word weighted automaton $\aut$ defines the function $\valueL{\aut}$ that assigns each word from $\Sigma^{\omega}$ its value $\valueL{\aut}(w)$.
This function is measurable for all the automata types we consider in this paper (see Remark~\ref{rem:measurability} below). 
Thus, this function can be interpreted as a random variable with respect to the probabilistic space we consider.
Hence, for a given automaton $\aut$ (over infinite words) and a Markov chain $\markov$, we consider the following quantities:
\medskip

\noindent\fbox{\parbox{0.96\textwidth}{
$\expected_{\markov}(\aut)$ --- the expected value of
the random variable $\valueL{\aut}$ w.r.t. the measure $\prob_{\markov}$.
\\
$\distrib_{\markov, \aut}(\const) = \prob_{\markov}(\{w \mid \valueL{\aut}(w) \leq \const \})$ --- the 
(cumulative) distribution function of  $\valueL{\aut}$ w.r.t. the measure $\prob_{\markov}$.
}}\medskip

In the finite words case, the expected value $\expected_{\markov^T}$ and the distribution $\distrib_{\markov^T, \aut}$ are defined in the same manner.

\begin{nremark}[Bounds on the expected value and the distribution]
Both quantities can be easily bounded: the value of the distribution function $\distrib_{\markov, \aut}$ is always between $0$ and $1$. 

For a $\flimavg$-automaton $\aut$, we have $\expected_{\markov}(\aut) \in [\min_\aut, \max_\aut] \cup \set{\infty}$, where $\min_\aut$ and $\max_\aut$ denote the minimal and the maximal weight of $\aut$ and $\expected_{\markov}(\aut) = \infty$ if and only if the probability of the  set of words with no accepting runs in $\aut$ is positive.
{Note that we consider no $\omega$-accepting conditions, and hence all infinite runs of $\flimavg$-automata are accepting, but there can be infinite words, on which a given $\flimavg$-automaton  has no infinite runs. }

For a $\fsum$-automaton $\aut$, we have $\expected_{\markov^T}(\aut) \in [L_{\markov^T} \cdot \min_\aut, L_{\markov^T} \cdot \max_\aut] \cup \set{\infty}$, where $L_{\markov^T}$ is the expected length of a word generated by $\markov^T$ 
(it can be computed in a standard way~\cite[Section 11.2]{Grinstead12}) 
and, as above, $\expected_{\markov^T}(\aut) = \infty$ if and only if there is a finite word $w$ generated by $\markov^T$ with non-zero probability such that $\aut$ has no accepting runs on $w$.

We show in Section~\ref{sec:irrational} that the distribution and expected value may be irrational, even for integer weights and uniform distributions.
\end{nremark}

\begin{nremark}[Measurability of functions represented by automata]
\label{rem:measurability}
For automata on finite words, $\finf$-automata and $\fsup$-automata, measurability of $\valueL{\aut}$ is straightforward.
To show that $\valueL{\aut}(w) \colon \Sigma^\omega \mapsto \R$ is measurable for any non-deterministic $\flimavg$-automaton $\aut$, it suffices to show that for every $x \in \R$, the preimage $\valueL{\aut}^{-1}(-\infty,x]$ is measurable.
Let  $Q$ be the set of states of $\aut$. We can define a subset $A_x \subseteq \Sigma^{\omega} \times Q^\omega$ of the pairs, 
the word and the run on it, where the value of the run is less than or equal to $x$.
We show that $A_x$ is Borel. For $p \in \N$, let $B_x^p$ be the subset of $\Sigma^{\omega} \times Q^\omega$ 
of pairs $(w, \pi)$
such that up to position $p$ the sequence $\pi$ is a run on $w$ and the average of weights up to $p$ is at most $x$.
Observe that $B_x^p$ is an open set and $A_x$ is equal to $\bigcap_{\epsilon \in \Q^+} \bigcup_{p_0 \in \N} \bigcap_{p \geq p_0} B_{x+\epsilon}^p$, i.e., 
$A_x$ consists of pairs $(w, \pi)$ satisfying that for every $\epsilon\in\Q^+$ there exists $p_0$ such that for all $p\geq p_0$ the average
weight of $\pi$ at $p$ does not exceed $x+\epsilon$ and $\pi$ is a run on $w$ (each finite prefix is a run).  
Finally, $\valueL{\aut}^{-1}(-\infty,x]$ is the projection of $A_x$ on the first component $\Sigma^{\omega}$. 
The projection of a Borel set is an \emph{analytic set}, which is measurable~\cite{kechris}. 
Thus, $\valueL{\aut}$ defined by  a non-deterministic $\flimavg$-automaton is measurable.

The above proof of measurability requires some knowledge of descriptive set theory. 
We will give a direct proof of measurability of $\valueL{\aut}$ in the paper (Theorem~\ref{th:approximation-limavg}).
\end{nremark}

\subsection{Computational questions} We consider the following basic computational questions:
\medskip

\noindent\fbox{\parbox{0.96\textwidth}{
    \emph{The expected value question}: Given an $f$-automaton $\aut$ and a  (terminating) Markov chain $\markov$, compute  $\expected_{\markov}(\aut)$.
    
\emph{The distribution question}: Given an $f$-automaton $\aut$, a (terminating) Markov chain $\markov$ and a threshold $\const \in \Q$, compute $\distrib_{\markov, \aut}(\const)$. 
}}\medskip

Each of the above questions have its decision variant (useful for lower bounds), where instead of computing the value we ask whether the value is less than a given threshold $t$. 

The above questions have their approximate variants:\medskip

\noindent\fbox{\parbox{0.96\textwidth}{
\emph{The approximate expected value question}:
Given an $f$-automaton $\aut$, a (terminating) Markov chain $\markov$, $\epsilon \in \Q^+$, compute a number $y \in \Q$ such that $|y - \expected_{\markov}(\aut)| \leq \epsilon$.

\emph{The approximate distribution question}: 
Given an $f$-automaton $\aut$, a (terminating) Markov chain $\markov$, a threshold $\const \in \Q$ and 
$\epsilon \in \Q^+$
compute a number $y \in \Q$ which belongs to $[\distrib_{\markov, \aut}(\const-\epsilon)-\epsilon, \distrib_{\markov, \aut}(\const+\epsilon)+\epsilon]$.
}}\medskip

\begin{nremark}
The notion of approximation for the distribution question is based on the Skorokhod metric~\cite{billingsley2013convergence}. 
Let us compare here this notion with two possible alternatives: the \emph{inside approximation}, where 
$y$ belongs to $[\distrib_{\markov, \aut}(\const-\epsilon), \distrib_{\markov, \aut}(\const+\epsilon)]$,
and the \emph{outside approximation}, where
$y$ belongs to $[\distrib_{\markov, \aut}(\const)-\epsilon, \distrib_{\markov, \aut}(\const)+\epsilon]$.

The outside approximation is reasonable for $\fsum$-automata, where the exact value of the probability is hard to compute, but for the $\flimavg$-automata its complexity is the same as computing the exact value 
(because the latter is difficult already for automata which return the same value for almost all words, as shown in Remark~\ref{r:ultimatelyperiodic}). 
For the inside approximation, it is the other way round: for $\fsum$-automata it makes little sense as the problem is undecidable even for automata returning integer values, but for $\flimavg$-automata it is a reasonable definition as the returned values can be irrational.

We chose a definition that works for both types of automata. 
However, the results we present can be easily adjusted to work in the case of the outside approximation for $\fsum$-automata and in the case of the inside approximation for the $\flimavg$-automata.
\end{nremark}

\section{Basic properties}

Consider an $f$-automaton $\aut$, a Markov chain $\markov$ and a set of words $X$. We denote by $\expected_{\markov}(\aut \mid X)$ the expected value of $\aut$ w.r.t. $\markov$ restricted only to words in the set $X$ (see \cite{feller}).
The following says that we can disregard a set of words with probability $0$ (e.g. containing only some of the letters under uniform distribution) while computing the expected value.

\begin{fact}\label{f:equal-expected}
If $\prob(X)=1$ then $\expected_\markov(\aut) = \expected_\markov(\aut \mid X)$.
\end{fact}

The proof is rather straightforward; the only interesting case is when there are some words not in $X$  with infinite values. 
But for all the functions we consider, one can show that in this case there is a set of words with infinite value that has a non-zero probability, and therefore $\expected_\markov(\aut) = \expected_\markov(\aut \mid X)=\infty$.

One corollary of Fact \ref{f:equal-expected} is that if $\markov$ is, for example, uniform, then because the set $Y$ of ultimately-periodic words (i.e., words of the form $vw^\omega$) is countable and hence has probability $0$, 
we have $\expected_\markov(\aut) = \expected_\markov(\aut \mid \Sigma^\omega \setminus Y)$. 
This suggests that the values of ultimately-periodic words might not be representative for an automaton. 
We exemplify this in Remark~\ref{r:ultimatelyperiodic}, where we show an automaton whose value is irrational for almost all words, yet rational for all ultimately-periodic words.

\subsection{Example of computing expected value by hand}\label{s:example}
Consider a $\flimavg$-automaton $\aut$ and a Markov chain $\markov$ depicted in Figure~\ref{fig:aut}. We encourage the reader to take a moment to study this automaton and try to figure out its expected value.

The idea behind $\aut$ is as follows. 
Assume that $\aut$ is in a state $q_l$ for some $l \in \{a,b\}$. Then, it reads a word up to the first occurrence of a subword $ba$, where it has a possibility to go to $q_x$ and then to non-deterministically  choose $q_a$ or $q_b$ as the next state. 
Since going to $q_x$ and back to $q_l$ costs the same as staying in $q_l$, we will assume that the automaton always goes to $q_x$ in such a case.
When an automaton is in the state $q_x$ and has to read a word $w=a^jb^k$, then the average cost of a run on $w$ is $\frac{j}{j+k}$ if the run goes to $q_b$ and $\frac{k}{j+k}$ otherwise. 
So the run with the lowest value is the one that goes to $q_a$ if $j>k$ and $q_b$ otherwise.

To compute the expected value of the automaton, we focus on the set $X$ of words $w$ such that for each positive $n \in \N$ there are only finitely many prefixes of $w$ of the form $w'a^jb^k$ such that $\frac{j+k}{|w'|+j+k} \geq \frac{1}{n}$. Notice that this means that $w$ contains infinitely many $a$ and infinitely many $b$. 
It can be proved in a standard manner that $\prob_\markov(X)=1$. 

Let $w \in X$ be a random event, which is a word generated by $\markov$. 
Since $w$ contains infinitely many letters $a$ and $b$, it can be partitioned in the following way.
Let $w=w_1w_2w_3 \dots$ be a partition of $w$ such that each $w_i$ for $i>0$ is of the form $a^jb^k$ for $j\geq 0,  k>0$, and for $i>1$ we also have $j>0$. 
For example, the partition of $w=baaabbbaabbbaba\dots$ is such that $w_1=b$, $w_2=aaabbb$, $w_3=aabbb$, $w_4=ab$, \dots. Let $s_i=|w_1w_2 \dots w_i|$.

We now define a run $\pi_w$ on $w$ as follows:
\[
q^w_1 \dots q^w_1 q_x q^w_2\dots q^w_2q_xq^w_3 \dots q^w_3 q_x q^w_4 \dots
\]
where the length of each block of $q_i$ is $|w_i|-1$, $q^w_0=q_a$ and 
$q^w_i=q_a$ if $w_i=a^jb^k$ for some $j>k$ and $q^w_i$=$q_b$ otherwise.
It can be shown by a careful consideration of all possible runs that this run's value is the infimum of values of all the runs on this word. 

\begin{lemma}\label{l:best-run}
For every $w \in X$ we have $\valueL{\aut}(w) = \flimavg(\pi_w)$.
\end{lemma}
\begin{proof}
We show that for every accepting run $\pi$ on $w \in X$ we have $\flimavg(\pi_w) \leq \flimavg(\pi)$. It follows that $\valueL{\aut}(w) = \flimavg(\pi_w)$.

Consider  a run $\pi$ of $\aut$ on $w$. 
The cost of a run over $w_i=a^jb^k$ is at least $min(j, k)-1$, which is reached by $\pi_w$, therefore
for every $i\in \N$ we have
\begin{equation}\label{e:property}
\favg{\pi_w[0,s_i]} \leq \favg{\pi[0,s_i]}.
\end{equation}

It may happen, however, that for some $p$, the value of $\favg{\pi[0,p]}$ is less than $\favg{\pi_w[0,p]}$; for example, for a word starting with $baaabbbb$, we have 
$\pi_w[0,4]=q_aq_xq_bq_bq_b$ and $\favg{\pi_w[0,4]}$ is $\frac{1}{2}$, but for a run $\pi'=q_aq_xq_aq_aq_a\dots$ we have $\favg{\pi'[0,4]}=0$. 
For arbitrary words, a run that never visits $q_b$ may have a better value. 
We show, however, that for words from $X$ this is not the case.

We show that for any position $p$ such that $s_i<p<s_{i+1}$, 
\begin{equation}\label{e:propertyTwo}
\favg{\pi_w[0,p]} \leq \favg{\pi[0,s_i]} + \frac{p-s_i}{p}
\end{equation}

Observe that 
\[ 
\begin{split}
\favg{\pi_w[0,p]}
= \frac{\fsum(\pi_w[0,p])}{p} &\leq \frac{\fsum(\pi_w[0,s_i])}{s_i} + \frac{\fsum(\pi_w[s_i,p])}{p} \\ &= \favg{\pi_w[0, s_i]} + \frac{\fsum(\pi_w[s_i,p])}{p}.
\end{split}
\] 
By \eqref{e:property} and the fact that the weights of the automaton do not exceed 1, 
we obtain 
\[ 
\favg{\pi_w[0, s_i]} + \frac{\fsum(\pi_w[s_i,p])}{p} \leq 
\favg{\pi[0, s_i]} + \frac{p-s_i}{p},
\] thus \eqref{e:propertyTwo}.

Assume $n \in \N$. By the definition of $X$, there can be only finitely many prefixes of $w$ of the form $w'a^jb^k$ where $\frac{j+k}{|w'|+j+k} \geq \frac{1}{n}$, so 
\(
\favg{\pi_w[0,p]} \geq \favg{\pi[0,s_i]} + \frac{1}{n}\)
may hold only for finitely many $p$.
Therefore, $\flimavg(\pi_w) \leq \flimavg(\pi) + \frac{1}{n}$ for every $n$, so $\flimavg(\pi_w) \leq \flimavg(\pi)$.
\end{proof}

By Fact~\ref{f:equal-expected} and Lemma~\ref{l:best-run}, it remains to compute the expected value of $\flimavg(\set{\pi_w \mid w \in X})$. 
As the expected value of the sum is the sum of expected values, we can state that 
\[\expected_\markov(\flimavg(\set{\pi_w \mid w \in X}))
=
\limsup\limits_{s \rightarrow \infty} \frac{1}{s} \cdot
\sum_{i=1}^{s} \expected_\markov\left(\set{(\cost(\pi_w))[i] \mid w \in X}\right)
\]

It remains to compute $\expected_\markov((\cost(\pi_w))[i])$.
If $i$ is large enough (and since the expected value does not depend on a finite number of values, we assume that it is),
the letter $\pi_w[i]$ is in some block $w_s=a^jb^k$. There are $j+k$ possible letters in this block, and the probability that the letter $\pi_w[i]$ is an $i$th letter in such a block is $2^{-(j+k+2)}$ (``+2'', because the block has to be maximal, so we need to include the letters before the block and after the block). So the probability that a letter is in a block $a^jb^k$ is $\frac{j+k}{2^{j+k+2}}$. The average cost of a such a letter is $\frac{\min(j, k)}{j+k}$, as there are $j+k$ letters in this block and the block contributes $\min(j, k)$ to the sum.

It can be analytically checked that 
\[
\sum_{j=1}^\infty 
\sum_{k=1}^\infty 
\frac{j+k}{2^{j+k+2}} \cdot \frac{\min(j, k)}{j+k}
=
\sum_{j=1}^\infty 
\sum_{k=1}^\infty 
\frac{\min(j, k)}{2^{j+k+2}} 
= \frac{1}{3}
\]
We can conclude that \(\expected_\markov(\flimavg(\pi_w))=\frac{1}{3}\) and, by Lemma \ref{l:best-run}, $\expected_\markov(\aut)=\frac{1}{3}$.

The bottom line is that even for such a simple automaton with only one strongly connected component consisting of three states (and two of them being symmetrical), the analysis is complicated. 
On the other hand, we conducted a simple Monte Carlo experiment in which we computed the value of this automaton on 10000 random words of length $2^{22}$ generated by $\markov$, and observed that the obtained values are in the interval $[0.3283, 0.3382]$,
with the average of $0.33336$, which is a good approximation of the expected value $0.(3)$. 
This foreshadows our results for $\flimavg$-automata: we show that computing the expected value is, in general, impossible, but it is possible to approximate it with arbitrary precision. 
Furthermore, the small variation of the results is not accidental -- we show that for strongly-connected $\flimavg$-automata, almost all words have the same value (which is equal to the expected value).

\subsection{Irrationality of the distribution and the expected value}\label{sec:irrational}
We show that the exact values in the probabilistic questions for $\fsum$-automata and $\flimavg$-automata may be (strongly) irrational.
More precisely, we show that for the $\fsum$-automaton depicted in Figure~\ref{fig:irrational}, the distribution $\distrib_{\aut}(-1)$ is transcendental, i.e., it is irrational and, unlike for instance $\sqrt{2}$, there is no
 polynomial with integer coefficients whose one of the roots is $\distrib_{\aut}(-1)$.
For the expected value, we construct  an automaton $\aut'$ such that $\expected(\aut)- \expected(\aut') = 1 - \distrib_{\aut}(-1)$ is transcendental. 
Therefore, one of $\expected(\aut)$, $\expected(\aut')$ is transcendental. 
Furthermore, we modify $\aut$ and $\aut'$ to show that there exists $\flimavg$-automaton $\autInf$ whose expected value is transcendental and the value $\const$ such that $\prob(\{w \mid \valueL{\autInf}(w) = \const) = 1$ is transcendental.
It follows that the minimal $\const$ such that $\distrib_{\autInf}(\const) = 1$ is transcendental.

\begin{theorem}[Irrational values]
\label{th:irrational}
The following conditions hold:
\begin{enumerate}
\item There exists a $\fsum$-automaton whose
distribution and expected value w.r.t. the uniform distribution are transcendental.  
\item There exists a $\flimavg$-automaton such that the expected value and
the value of almost all words w.r.t. the uniform distribution are transcendental.
\end{enumerate}
\end{theorem}
\begin{proof}
We assume that the distribution of words is uniform. 
In the infinite case, this means that the Markov chain contains a single state where it loops over any letter with probability $\frac{1}{|\Sigma|}$, where $\Sigma$ is the alphabet. 
In the finite case, this amounts to a terminating Markov chain with one regular state and one terminating state; it loops over any letter in the non-terminating state with probability $\frac{1}{|\Sigma|+1}$ or 
go to the terminating state over $\epsilon$ with probability $\frac{1}{|\Sigma|+1}$.
Below we omit the Markov chain as it is fixed (for a given alphabet).

We define a $\fsum$-automaton $\aut$ (Figure~\ref{fig:irrational}) over the alphabet $\Sigma = \set{a, \# }$ such that
$\aut(w) = 0$ if $w = a \# a^4 \# \ldots \# a^{4^n}$ and $\aut(w) \leq -1$ otherwise. 
Such an automaton basically picks a block with an inconsistency and verifies it. 
For example, if $w$ contains a block $\# a^i \# a^j \#$, the automaton $\aut$ first assigns $-4$ to each letter $a$ 
and upon $\#$ it switches to the mode in which it assigns $1$ to each letter $a$. Then, $\aut$ returns the value $j - 4\cdot i$. Similarly, 
we can encode the run that returns the value  $4\cdot i - j$. Therefore, all the runs return $0$ if and only if each block of $a$'s is four times as long as the previous block. 
Finally, $\aut$ checks whether the first block of $a$'s has length $1$ and returns $-1$ otherwise.

Let $\gamma$ be the probability that a word is of the form $a \# a^4 \# \ldots \# a^{4^n}$.
Such a word has length $l_n = \frac{4^{n+1} -1}{3}+n$ and its probability is 
${ 3^{-(l_n+1)} }$ (as the probability of any given word with $m$ letters over a two-letters alphabet is $3^{-(m+1)}$).
Therefore $\gamma$ is equal to $\sum_{n=0}^{\infty} { 3^{-(l_n+1)} }$. 
Observe that $\gamma$ written in base $3$ has arbitrary long sequences of $0$'s and hence its representation is acyclic. Thus, $\gamma$ is irrational.

Due to Roth's Theorem~\cite{roth}, if $\alpha \in \R$ is algebraic but irrational, then there are only finitely many pairs $(p,q)$ such that
$|\alpha - \frac{p}{q}| \leq \frac{1}{q^3}$. We show that there are infinitely many such pairs for $\gamma$ and hence it is transcendental.
Consider $i \in \N$ and let $p_i, q_i \in \N$ be such that $q_i = 3^{-(l_i+1)}$ and
$\frac{p_i}{q_i} = \sum_{n=0}^{i} { 3^{-(l_n+1)}}$. 
Then, \[ 0 < \gamma - \frac{p_i}{q_i} < 2\cdot 3^{-(l_{i+1}+1)} \]
Observe that for $i>1$ we have $l_{i+1} >  3 (l_{i}+1)$ and hence
\[
 \gamma - \frac{p_i}{q_i} < 2\cdot 3^{-(l_{i+1}+1)} <  \frac{2}{3} 3^{-3(l_i+1)} < \frac{1}{q_i^3}.
\]
Therefore, 
$\gamma$ is transcendental.

Observe that $\gamma = 1 - \distrib_{\aut}(-1)$. Therefore, $\distrib_{\aut}(-1)$ is transcendental.
For the expected value, we construct $\aut'$ such that for every word $w$ we have $\valueL{\aut'}(w) = min(\valueL{\aut}(w),-1)$. 
This can be done by adding to $\aut$ an additional initial state $q_0$, which starts an automaton that assigns to all words value $-1$.
Observe that $\aut$ and $\aut'$ differ only on words $w$ of the form $a \# a^4 \# \ldots \# a^{4^n}$, where $\aut(w) = 0$ and $\aut'(w) = -1$.
On all other words, both automata return the same values. Therefore,  $\expected(\aut) - \expected(\aut') = \gamma$.
It follows that at least one of the values $\expected(\aut)$, $\expected(\aut')$ is transcendental.   

The same construction works for $\flimavg$-automata. 
We take $\aut$ defined as above and convert it to a $\flimavg$-automaton $\autInf$ over $\Sigma'  = \Sigma \cup \{\$\}$, where the fresh letter $\$$ resets the automaton, 
i.e., $\autInf$ has transitions labeled by $\$$ from any final state of $\aut$ to any of its initial states. We apply the same construction to $\aut'$ defined as above and denote the resulting automaton by $\autInf'$.
Observe that  $\expected(\autInf) = \expected(\aut)$ (resp., $\expected(\autInf') = \expected(\aut')$.
To see that, consider random variables $X_1, X_2, \ldots$ defined on $\Sigma^{\omega}$, where $X_i(w)$ is the average value $\frac{1}{|u|}\valueL{\aut}(u)$ of the $i$-th block $\$u\$$ in $w$, i.e., 
$w =u_1 \$ u_2 \$ \ldots \$u_i \$ \ldots$, all $u_j$ are from $\Sigma^*$ and $u = u_i$.
Observe that $X_1, X_2, \ldots$ are independent and identically distributed random variables and hence with probability $1$ we have
\[
 \liminf\limits_{s \rightarrow \infty} \frac{1}{s} (X_1 + \ldots + X_s) =  \limsup\limits_{s \rightarrow \infty} \frac{1}{s} (X_1 + \ldots + X_s) = \expected(X_i) = \expected(\aut)
\]
Therefore, with probability $1$ over words $w$ we have $\valueL{\autInf}(w) = \expected(\aut)$. It follows that
$\expected(\autInf) = \expected(\aut)$ and the minimal $\const$ such that 
$\distrib_{\autInf}(\const) = 1$ equals $\expected(\aut)$. 
Similarly, $\expected(\autInf') = \expected(\aut')$ and the minimal $\const$ such that $\distrib_{\autInf'}(\const) = 1$ equals $\expected(\aut')$. 
Therefore, for one of automata $\autInf, \autInf'$, the value of almost all words and  the expected value are transcendental.
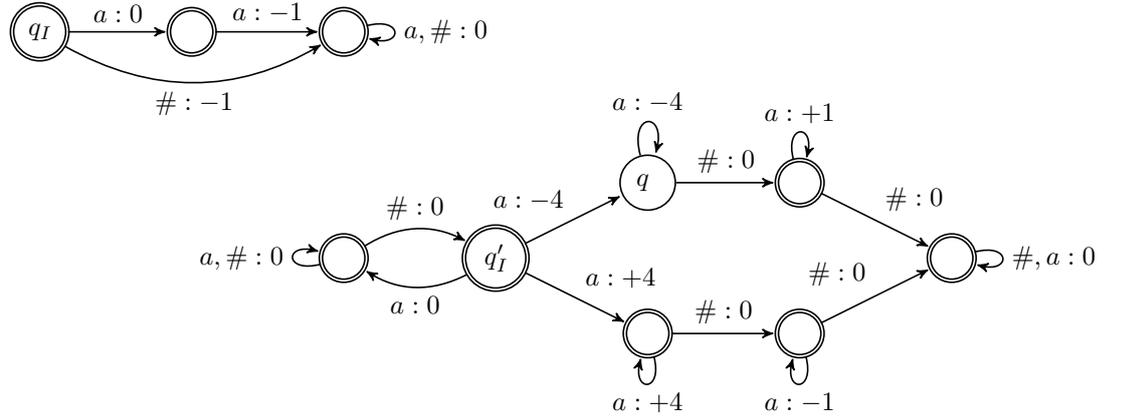
\begin{figure}
{
\centering
\begin{tikzpicture}[->,>=stealth',shorten >=1pt,auto,node distance=2.0cm,semithick]
  \tikzstyle{every state}=[fill=white,draw=black,text=black,minimum size=1.8cm]
  \tikzset{
   datanode/.style = {
    draw,
    circle,
    text width=0.3cm,
    inner ysep = +0.4em},
    labelnode/.style = {
    draw, 
    rounded corners,
    align=center,
    fill=white,
    minimum size=8.8cm}
  }

\node at (-6,0) {}; 

\foreach \lab\lX/\lY/\name/\label in 
         {Aa/-4/0/$q_I$/double,
          Ab/-2/0//double,
          Ac/0/0//double,
Ca/0/-3//double,
          Cb/2/-3/$q_I'$/double,
          Cc/4/-2/$q$/draw,         
          Cd/6/-2//double,
          Ce/8/-3//double,
          Dc/4/-4//double,         
          Dd/6/-4//double}
{
  
  \node[datanode,\label] (\lab) at (\lX,\lY)  
  {
   \name
  };
  
}

\draw  (Aa) edge node {$a:0$} (Ab);
\draw  (Aa) edge[bend right] node[below] {$\#:-1$} (Ac);
\draw  (Ab) edge node {$a:-1$} (Ac);
\draw  (Ac) edge[loop right] node {$a, \#: 0$} (Ac);

\draw  (Ca) edge[loop left] node {$a, \#:0$} (Ca);
\draw  (Ca) edge[bend left] node {$\#:0$} (Cb);
\draw  (Cb) edge[bend left] node {$a:0$} (Ca);
\draw  (Cb) edge node {$a:-4$} (Cc);
\draw  (Cc) edge[loop above] node {$a:-4$} (Cc);
\draw  (Cc) edge node {$\#:0$} (Cd);
\draw  (Cd) edge[loop above] node {$a:+1$} (Cd);
\draw  (Cd) edge node {$\#:0$} (Ce);
\draw  (Ce) edge[loop right] node {$\#, a:0$} (Ce);

\draw  (Cb) edge node {$a:+4$} (Dc);
\draw  (Dc) edge[loop below] node {$a:+4$} (Dc);
\draw  (Dc) edge node {$\#:0$} (Dd);
\draw  (Dd) edge[loop below] node {$a:-1$} (Dd);
\draw  (Dd) edge node {$\#:0$} (Ce);

\end{tikzpicture}
}
 \caption{The automaton $\aut$ from Section \ref{sec:irrational}. States $q_I$ and $q_I'$ are initial and states  but $q$ are accepting. Any word that starts with $\#$ or $aa$ has the value at most $-1$ because of a run that starts in $q_I$. 
For all other words, the runs starting in $q_I$ have value $-1$. 
The accepting runs starting in $q_I'$ have negative value only if the input word contains a (maximal) subword $a^i\#a^j$ such that $j \neq 2i$. }\label{fig:irrational}
\end{figure}
\end{proof}
 
\section{The exact value problems}
\label{s:exact}
In this section we consider the probabilistic questions for non-deterministic $\fsum$-automata and $\flimavg$-automata, i.e., the problems of computing the exact values of the expected value $\expected_{\markov}(\aut)$ and the distribution $\distrib_{\markov, \aut}(\const)$  w.r.t. a Markov chain $\markov$ and an $f$-automaton $\aut$.
The answers to these problems are related values may be irrational (Theorem~\ref{th:irrational}), but 
 one can perhaps argue that there might be some representation of irrational numbers that can be employed to avoid this problem.
We prove that this is not the case by showing that computing the exact value to any representation with 
decidable equality of two numbers is impossible.

\begin{theorem}\label{th:limavg-undecidable}
The following conditions hold:
\begin{enumerate}
\item The expected value and the distribution of (non-deterministic) $\fsum$-automata are uncomputable even for the uniform probability measure. 
\item The expected value and the value of almost all words (if it exists) of (non-deterministic)  $\flimavg$-automata are uncomputable even for the uniform probability measure. 
\end{enumerate}
\end{theorem}

\begin{proof}
The proof is by a (Turing) reduction from the quantitative universality problem for $\fsum$-automata, which is undecidable~\cite{Krob94,AlmagorBK11}: \medskip

\noindent\fbox{\parbox{0.96\textwidth}{\emph{The quantitative universality problem for $\fsum$-automata}: Given a $\fsum$-automaton with weights $-1$, $0$ and $1$, decide whether for all words $w$ we have $\valueL{\aut}(w) \leq 0$.
}}\medskip

We first discuss reductions to the probabilistic problems for $\fsum$-automata.
Consider an instance of the quantitative universality problem, which is a $\fsum$-automaton $\aut$. 
If there is a word $w$ with the value greater than $0$, then due to uniformity of the probability measure we have $\prob(w)>0$, and thus $\distrib_{\aut}(0) < 1$. 
Otherwise, clearly $\distrib_{\aut}(0) = 1$. 
Therefore, solving the universality problem amounts to computing whether the $\distrib_{\aut}(0) = 1$, and thus the latter problem is undecidable.
For the expected value, we construct a $\fsum$-automaton $\aut'$ 
such that for every word $w$ we have $\valueL{\aut'}(w) = min(\valueL{\aut}(w),0)$. 
Observe that $\expected(\aut) = \expected(\aut')$ if and only if for every word $w$ we have   $\valueL{\aut}(w) \leq 0$, i.e., the answer to the universality problem is YES. 
Therefore, there is no Turing machine, which given a $\fsum$-automaton $\aut$ computes  $\expected(\aut)$ (in any representation allowing for effective equality testing).

For the $\flimavg$ case, we construct a $\flimavg$-automaton $\autInf$ from the $\fsum$-automaton $\aut$, by connecting all accepting states (of $\aut$) with all initial states by transitions of weight $0$
labeled by an auxiliary letter $\#$. 
We construct $\autInf'$ from $\aut'$ in the same way.
The automata $\autInf, \autInf'$ have been constructed from $\aut$ and respectively $\aut'$ as in the proof of Theorem~\ref{th:irrational}, and the virtually the same argument shows that
for almost all words $w$ (i.e., with the probability $1$) we have  $\valueL{\autInf}(w) = \expected{\aut}$ (resp., $\valueL{\autInf'}(w) = \expected(\aut')$).
Therefore, $\expected(\autInf) = \expected(\autInf')$ if and only if for every finite word $u$ we have   $\valueL{\aut}(u) \leq 0$.
In consequence, there is no Turing machine computing the expected value of a given $\flimavg$-automaton.
Furthermore, since $\autInf$ (resp., $\autInf'$) returns $\expected(\autInf)$ (resp., $\expected(\autInf')$)  on almost all words, there is no Turing machine computing
 the value of almost all words of a given (non-deterministic)  $\flimavg$-automaton.
\end{proof}

\subsection{Extrema automata}
We discuss the distribution problem for $\fmin$-, $\fmax$-, $\finf$- and $\fsup$-automata, where
$\fmin$ and $\fmax$ return the minimal and respectively the maximal element of a finite sequence, and 
$\finf$ and $\fsup$ return the minimal and respectively the maximal element of an infinite sequence.
The expected value of an automaton can be easily computed based on the distribution as there are only finitely many possible values of a run (each possible value is a label of some transition).

\begin{theorem}
\label{th:extrema}
For $\fmin$-, $\fmax$-, $\finf$- and $\fsup$-automata $\aut$ and a Markov chain $\markov$, 
the expected value and
the distribution problems can be solved in exponential time in $|\aut|$ and polynomial time in $|\markov|$.
\end{theorem}
\begin{proof}
We discuss the case of $f = \finf$ as the other cases are similar. 
Consider an $\finf$-automaton $\aut$. 
Observe that every value returned by $\aut$ is one of its weights.
For each weight $x$ of $\aut$, we construct a (non-deterministic) $\omega$-automaton $\aut_x$ that accepts only words of value greater than $x$, i.e., 
$\valueL{\aut_x} = \{ w \mid \valueL{\aut}(w) > x \}$. To construct $\aut_x$, we take $\aut$, remove the transitions of weight less or equal to $x$, and drop all the weights. 
Therefore, the set of words with the value greater than $x$ is regular, and hence it is measurable and we can compute its probability $p_x$ by computing the probability of $\valueL{\aut_x}$.
The probability of an $\omega$-regular language given by a non-deterministic $\omega$-automaton (without acceptance conditions) can be computed in exponential time in the size of the automaton and
polynomial time in the Markov chain defining the probability distribution~\cite[Chapter 10.3]{BaierBook}. 
It follows that $p_x$ can be computed in exponential time in $|\aut|$ and polynomial time in $|\markov|$.

Observe that $p_x = 1 - \distrib_{\markov,\aut}(x)$ and hence we can compute the distribution question $\distrib_{\markov,\aut}(\const)$ by computing $1 - p_x$ for the maximal weight $x$ that does not exceed $\lambda$.
For the expected value, let $x_1, \ldots, x_k$ be all weights of $\aut$ listed in the ascending order. 
Let $p_0 = 1$.
Observe that for all $i \in \set{1, \ldots, k}$ we have $p_{x_{i-1}} - p_{x_i} = \prob_{\markov}(\set{w \mid \valueL{\aut}(w) = x_i})$ is the probability of the set of words of the value $x_i$.
Therefore, $\expected_{\markov}(\aut) = \sum_{i=1}^{k} (p_{x_{i-1}} - p_{x_i}) \cdot x_i$ and hence the expected value can be computed in exponential time in $|\aut|$ and polynomial time in $|\markov|$.
\end{proof}
 
\section{The approximation problems}
\label{s:approx}
\newcommand{\fin}{\mathrm{fin}}
\newcommand{\gKA}{g[{k}]}

\newcommand{\geKA}{g[{2^k}]}
\newcommand{\exactMarkov}{M[k]}
\newcommand{\compMarkov}{N[k]}

\newcommand{\apxMarkov}{\markov^{\approx}}
\newcommand{\optRun}{\xi_o}
\newcommand{\folRun}{\pi_f}
\newcommand{\diffMarkov}{\markov^{diff}}

We start the discussion on the approximation problems by showing a hardness result that holds for a wide range of value functions.
We say that a function is $0$-preserving if its value is $0$ whenever the input consists only of $0$s. 
The functions $\fsum$, $\flimavg$, $\fmin$, $\fmax$, $\finf$, $\fsup$ and virtually all the functions from the literature~\cite{quantitativelanguages} are $0$-preserving. 
The hardness result follows from the fact that accepted words have finite values, which we can force to be $0$, while words without accepting runs have infinite values. 

The answers in the approximation problems are numbers and to study the lower bounds, we consider their decision variants, called the \emph{separation problems}.
The \emph{expected separation problem} is a variant of the expected value problem, in which the input is enriched with numbers $a, b$ such  that $b-a>2\epsilon$ 
 and the instance is such that $\expected_{\markov}(\aut) \not \in [a, b]$ and the question is whether $\expected_{\markov}(\aut)<a$.
In the \emph{distribution separation problem}, the input is enriched with numbers $a,b,c,d$ such that $b-a >2\epsilon$ and $d-c >2\epsilon$, the instance satisfies 
for all $\lambda \in [c,d]$ we have $\distrib_{\markov,\aut}(\lambda) \not \in [a, b]$, and we ask whether $\distrib_{\markov,\aut}(\frac{c+d}{2})<a$.
Note that having an algorithm computing one of the approximate problems (for the distribution or the expected value), we can use it to decide the separation question. 
Conversely, using the separation problem as an oracle, we can perform binary search on the domain 
to solve the corresponding approximation problem in polynomial time.
\begin{theorem}
\label{th:all-pspace-hard}
The following conditions hold:
\begin{enumerate}
\item For any $0$-preserving function $f$, the expected separation problem for non-deterministic $f$-automata is $\PSPACE$-hard.
\item For any $0$-preserving function $f$ over infinite words, the distribution separation problem for non-deterministic $f$-automata over infinite words is $\PSPACE$-hard.
\end{enumerate}
\end{theorem}
\begin{proof}
The proof is via reduction from the universality question for non-deterministic (unweighted) finite-word automata, which is  $\PSPACE$-complete~\cite{HU79}.

\noindent\emph{The finite-word case}. 
We consider the uniform distribution over finite words.
Given a non-deterministic finite-word automaton $\aut$, we construct a finite-word $f$-automaton $\aut_{\fin}$ by labeling all transitions of $\aut$ with $0$.
Observe that if there exists a word which is not accepted by $\aut$ then the expected value of $\aut_{\fin}$ is $\infty$. 
Otherwise, all words have value $0$ and hence the expected value for $\aut_{\fin}$ is $0$.
The universality problem for $\aut$  reduces to the expected separation problem for $\aut_{\fin}$.

\noindent\emph{The infinite-word case}. 
We consider the uniform distribution over infinite words.
Given a non-deterministic finite-word automaton $\aut$, we construct an infinite-word $f$-automaton $\autInf$ in the following way.
We start with the automaton $\aut$. 
First, we extend the input alphabet with a fresh letter $\#$, which resets the automaton.
More precisely, we add transitions labeled by $\#$ between any final state of $\aut$ and any initial state of $\aut$.
Finally, we label all transitions with $0$. The resulting automaton is $\autInf$.

If there exists a finite word $u$ rejected by $\aut$, then for every infinite word $w$ containing the infix $\# u \#$ the automaton $\autInf$ has no infinite run and hence it assigns value $\infty$ to $w$.
Observer that the set of words containing $\# u \#$ has probability $1$ (for any finite word $u$). Therefore, if $\aut$ rejects some word, 
the expected value for $\autInf$ is $\infty$ and the distribution of $\autInf$ for any $\lambda \in \R$ is $0$.
Otherwise, if $\aut$ accepts all words, the expected value of $\autInf$  is $0$ and  the distribution of $\autInf$ for any $\lambda \geq 0$ is $1$.
The universality problem for $\aut$  reduces to the separation problems for $\autInf$.
\end{proof}

\Paragraph{Total automata} Theorem~\ref{th:all-pspace-hard} gives us a general hardness result, which is due to accepting conditions rather than values returned by weighted automata.
In the following, we focus on weights and we assume that weighted automata are \emph{total}, i.e., they accept all words 
(resp., almost all words in the infinite-word case). 
For $\fsum$-automata under the totality assumption, the approximate probabilistic questions become \sharpP-complete.
We additionally show  that the approximate distribution question for $\fsum$-automata is in \sharpP{} regardless of totality assumption.

\begin{theorem}
\label{th:approximation}
The following conditions hold:
\begin{enumerate}
\item The approximate expected value and the approximate distribution questions for total non-deterministic total $\fsum$-automata are \sharpP-complete.
\item The approximate distribution question for non-deterministic $\fsum$-automata is \sharpP-complete.
\end{enumerate}
\end{theorem}
\begin{proof}
\noindent\emph{\sharpP-hardness}. 
Consider the problem of counting the number of satisfying assignment of a given 
propositional formula $\varphi$ in Conjunctive Normal Form (CNF)~\cite{valiant1979complexity,papadimitriou2003computational}. This problem is \sharpP-complete. We reduce it to
  the problem of approximation of the expected value for total $\fsum$-automata. 
Consider a formula $\varphi$ in CNF over $n$ variables. 
Let $\markov^T$ be a terminating Markov chain over $\{0,1\}$, which at each step produces $0$ and $1$ with equal probability $\frac{1}{3}$, and it terminates with probability $\frac{1}{3}$.
We define a total $\fsum$-automaton $\aut_{\varphi}$ such that it assigns $0$ to all words of length different than $n$. For words $u \in \set{0,1}^n$, 
the automaton $\aut_{\varphi}$ regards $u$ as an assignment for variables of $\varphi$;  $\aut_{\varphi}$ non-deterministically picks one clause of $\varphi$ and returns $1$ if that clause is satisfied and $0$ otherwise.
We can construct such $\aut_{\varphi}$ to have polynomial size in $|\varphi|$.
Observe that $\aut_{\varphi}(u) = 0$ if some clause of $\varphi$ is not satisfied by $u$, i.e., $\varphi$ is false under the assignment given by $u$.
Otherwise, if the assignment given by $u$ satisfies $\varphi$, then $\aut_{\varphi}(u) = 1$. 
It follows that the expected value of $\aut_{\varphi}$ equals ${3}^{-(n+1)} \cdot C$, where ${3}^{-(n+1)}$ is the probability of generating a word of length $n$ and 
$C$ is the number of variable assignments satisfying $\varphi$. 
Therefore, we can compute $C$ by computing the expected value of $\aut_{\varphi}$ with any $\epsilon$ less than $0.5 \cdot {3}^{-(n+1)}$.
Observe that the automaton $\aut_{\varphi}$ returns values $0$ and $1$ and hence 
the expected value $\expected_{\markov}(\aut_{\varphi}) = 1  - \distrib_{\markov, \aut_{\varphi}}(0)$, where $1  - \distrib_{\markov, \aut_{\varphi}}(0)$ is the probability that
$\aut_{\varphi}$ returns $1$.
\smallskip

\noindent\emph{Containment of the approximate distribution question in \sharpP}.
Consider a terminating Markov chain $\markov^T$, a $\fsum$-automaton $\aut$, and $\epsilon \in \Q^+$. 

Let $C$ be the smallest number such that every non-zero  probability in $\markov^T$ is at least $2^{-C}$. 
Such $C$ is polynomial in the input size. Consider $N=C \cdot \mathrm{len}(\epsilon)+1$ and let 
$\distrib_{\markov^T, \aut}(\const, N)$ be the distribution of $\aut$ over words up to length $N$, i.e., $\prob_{\markov^T}(\{w \mid |w| \leq N\ \wedge \valueL{\aut}(w) \leq \const\})$.
 We show that the distribution of $\aut$ and the distribution of $\aut$ over words up to length $N$ differ by less than $\frac{\epsilon}{2}$, i.e., that
\[
|\distrib_{\markov^T, \aut}(\const) - \distrib_{\markov^T, \aut}(\const,n)| \leq \frac{\epsilon}{2}.
\] 

To do so, let $p_n$, for $n \in \N$, be the probability that $\markov^T$ emits a word of the length greater than $n$. 
From any state of $\markov^T$, the probability of moving to a terminating state is at least $2^{-C}$. 
We can (very roughly) bound the probability of generating a word of length greater than $i$ ($p_{i})$ by $(1-2^{-C})^i$. 
This means that $p_n$ decreases exponentially with $n$.
Since $(1-\frac{1}{n})^n \leq \frac{1}{2}$ for all $n>1$, we obtained the desired inequality.

Let $K = (N+1)\cdot \log(|\Sigma|) \cdot \epsilon^{-1}+1$. 
We build a non-deterministic Turing machine $H_1$ such that on the input $\markov^T$, $\aut$, $\epsilon$, and $\lambda$ such that the number $c_A$ of accepting computations of $H_1$ satisfies the following:

\[
\Bigl\lvert \distrib_{\markov^T,\aut}(\lambda, n) - \frac{c_A}{2^K}\Bigl\rvert  \leq \frac{\epsilon}{2}.
\]

To $\epsilon$-approximate $\distrib_{\markov^T,\aut}(\lambda)$, we need to compute $c_A$ and divide it by $2^K$, which can be done in polynomial time.

The machine $H_1$ works as follows.
Given the input $\markov^T$, $\aut$, $\epsilon$, it non-deterministically generates a string $u\alpha$, where $u \in (\Sigma \cup \{\#\})^N$ is a word and $\alpha \in \set{0, 1}^K$ is a number written in binary.
The machine rejects unless $u$ is of the form $wv$, where $w\in \Sigma^*$ and $v\in\set{\#}^*$.
Then, the machine accepts if 
$\valueL{\aut}(w) \leq \lambda$
and 
$\alpha \leq 2^K \cdot \prob_{\markov^T}(w)$.
Therefore, provided that $H_1$ generates $w$ with $\valueL{\aut}(w) \leq \lambda$, the number of accepting computations $c_A^w$ equals $\lfloor 2^K \cdot \prob_{\markov^T}(w) \rfloor$. It follows that
$c_A^w$  divided by $2^K$ is a $2^{-K}$-approximation of $\prob_{\markov^T}(w)$, i.e., 
\[
\Bigl\lvert  \prob_{\markov^T}(w) - \frac{c_A^w}{2^K} \Bigl\rvert  < 2^{-K}.
\]   

The total number of accepting paths of $H_1$ is given by
\[
c_A = \sum_{w \colon |w| \leq N\ \wedge \valueL{\aut}(w) \leq \const} c_A^w.
\]
We estimate the difference between $\distrib_{\markov^T,\aut}(\lambda, n)$ and the value $\frac{c_A}{2^K}$:
\[
\Bigl\lvert \distrib_{\markov^T,\aut}(\lambda, n) -  \frac{c_A}{2^K} \Bigr\rvert  \leq 
\sum_{w \colon |w| \leq N\ \wedge \valueL{\aut}(w) \leq \const} \Bigr\lvert  \prob_{\markov^T}(w) - \frac{c_A^w}{2^K}\Bigr\rvert  \leq  |\Sigma|^{N+1} \cdot 2^{-K} 
< \frac{\epsilon}{2}.
\]

\noindent\emph{Containment of the approximate expected value question in \sharpP}
\newcommand{\expectedN}{M}
Assume that $\aut$ is total. 
For readability we assume that $\aut$ has only integer weights. If it does not, we can multiply all weights by least common multiple of all denominators of weights in $\aut$; this operation multiples the expected value by the same factor.

Recall that $C$ is the smallest number such that every non-zero probability in $\markov^T$ is at least $2^{-C}$.
Let $W$ be the maximal absolute value of weights in $\aut$ and
let $\expectedN = C \mathrm{len}(\epsilon)\cdot \log(C W) +1$ and $\expected_{\markov^T}(\aut, N)$ be the expected value of $\markov^T$ for words up to length $\expectedN$, i.e., computing only the finite sum from the definition of the expected value. 
We show that 
\[\bigl\lvert \expected_{\markov^T}(\aut)-\expected_{\markov^T}(\aut,\expectedN)\bigl\rvert  \leq \frac{\epsilon}{2}.\]
Recall that $p_n$ is the probability that $\markov^T$ emits a word of the length greater than $n$, and $p_{n} \leq (1-2^{-C})^n$.
Since $\aut$ is total, the value of every word $w$ is finite and it belongs to the interval $[-|w| \cdot W, |w| \cdot W]$.
The value of a word of the length bounded by $i$ is at most $i \cdot W$.
Therefore, the expected value of $\aut$ over words of grater than $k$ is bounded from above by 
\(
\sum_{i \geq k} p_{i} \cdot i \cdot W \leq W \cdot (1-  2^{-C})^k \cdot (k+1)
\).

W.l.o.g. we assume, that there are no transitions to the initial state in $\aut$.
Next, we transform $\aut$ to an automaton $\aut'$ that returns natural numbers on all words of length at most $\expectedN$ by adding $W \cdot \expectedN$ to every transition from the initial state.
Observe that $\expected_{\aut'} = \expected_{\aut} + W\cdot \expectedN$ and 
$D = 2\cdot W \cdot \expectedN$ is an upper bound on values returned by the automaton $\aut'$ on words of length at most $\expectedN$.

Finally, we construct a Turing machine $H_2$, similar to $H_1$.
Let $K = (D+1) \cdot (N+1)\cdot (|\Sigma|+1) \cdot \epsilon^{-1}+1$. 
$H_2$ non-deterministically chooses a word $u\alpha$, where 
$u \in (\Sigma \cup \{\#\})^N$ is a word and $\alpha \in \set{0, 1}^K$ is a number written in binary, and also non-deterministically picks a natural number $\beta \in [0, D]$.
The machine rejects unless $u$ is of the form $wv$, where $w\in \Sigma^*$ and $v\in\set{\#}^*$.
Then $H_2$ accepts if and only if $\valueL{\aut'}(w) \leq \beta$
and $\alpha \leq 2^K \cdot \prob_{\markov^T}(w)$. 
Then, provided that $H_2$ generates $w$, the number of accepting computations $c_A^w$ equals $\lfloor 2^K \cdot \prob_{\markov^T}(w) \cdot \valueL{\aut'}(w) \rfloor$.
Therefore, using estimates similar to the distribution case, we obtain the desired inequality
\[\bigl\lvert \expected_{\markov^T}(\aut',\expectedN) - \frac{c_A}{2^K}\bigl\rvert  \leq \frac{\epsilon}{2}.\]
Finally, we obtain that $\frac{c_A}{2^K} -WM$ is an $\epsilon$-approximation of $ \expected_{\markov^T}(\aut)$, i.e.,
\[\bigl\lvert \expected_{\markov^T}(\aut) - \big(\frac{c_A}{2^K} -WM\big)\bigl\rvert  \leq {\epsilon}.\]
\end{proof}

We show that the approximation problem for $\flimavg$-automata is $\PSPACE$-hard over the class of total automata.

\begin{theorem}
\label{th:approximation-sum}
The separation problems for non-deterministic total $\flimavg$-automata are $\PSPACE$-hard.
\end{theorem}
\begin{proof}
We consider the uniform distribution over infinite words.
Given a non-deterministic finite-word automaton $\aut$, we construct an infinite-word $\flimavg$-automaton $\autInf$ from $\aut$ in the following way.
We introduce an auxiliary symbol $\#$ and we add transitions labeled by $\#$ between any final state of $\aut$ and any initial state of $\aut$.
Then, we label all transitions of $\autInf$ with $0$.
Finally, we connect all non-accepting states of $\aut$ with an auxiliary state $q_{\mathrm{sink}}$, which is a sink state with all transitions of weight $1$.
The automaton $\autInf$ is total.

Observe that if $\aut$ is universal, then $\autInf$ has a run of value $0$ on every word. Otherwise, if $\aut$ rejects a word $w$, then
upon reading a subword $\# w \#$, the automaton $\autInf$ reaches $q_{\mathrm{sink}}$, i.e., the value of the whole word is $1$.
Almost all words contain an infix  $\# w \#$ and hence almost all words have value $1$. 
Therefore, the universality problem for $\aut$ reduces to the problem
deciding whether for almost all words $w$ we have  $\valueL{\autInf}(w) = 0$ or 
for almost all words $w$ we have $\valueL{\autInf}(w) = 1$? The latter problem reduces to the expected separation problem as well as the distribution separation problem for $\autInf$.
\end{proof}

\section{Approximating $\flimavg$-automata in exponential time}

In this section we develop algorithms for the approximate expected value and approximate distribution questions 
for (non-deterministic) $\flimavg$-automata. 
The presented algorithms work in exponential time in the size of the automaton, polynomial time in the size of the Markov chain and the precision. 

The case of $\flimavg$-automata is significantly more complex than the other cases and hence we present the algorithms in stages. 
First, we restrict our attention to \emph{recurrent} $\flimavg$-automata and the uniform distribution over infinite words. 
Recurrent automata are strongly connected with an appropriate set of initial states. 
We show that deterministic $\flimavg$-automata with bounded look-ahead approximate recurrent automata. 
Next, in Section~\ref{s:non-uniform} we extend this result to non-uniform measures given by Markov chains.
Finally, in Section~\ref{s:non-recurrent} we show the approximation algorithms for all (non-deterministic) $\flimavg$-automata and measures given by Markov chains.

\Paragraph{Recurrent automata}
Let $\aut = (\Sigma, Q, Q_0,\delta)$ be a non-deterministic $\flimavg$-automaton and $\widehat{\delta}$ be the extension of $\delta$
to all words $\Sigma^*$. The automaton $\aut$ is \emph{recurrent} if and only if the following conditions hold:
\begin{enumerate}[(1)]
\item for every state $q \in Q$ there is a finite word $u$ such that $\widehat{\delta}(q, u) = Q_0$ ($\widehat{\delta}$ is the transition relation extended to words), and 
\item for every set $S \subseteq Q$, if $\widehat{\delta}(Q_0, w) = S$ for some word $w$, then there is a finite word $u$ such that $\widehat{\delta}(S, u) = Q_0$. 
\end{enumerate}

Intuitively,  in recurrent automata $\aut$, if two runs deviate at some point, with high probability it is possible to synchronize them. 
More precisely, for almost all words $w$, if $\pi$ is a run on $w$, and $\rho$ is a finite run up to position $i$, then $\rho$ can be extended to an infinite run that eventually coincides with $\pi$.
Moreover, we show that with high probability, they synchronize within doubly-exponential number of steps in $|\aut|$ (Lemma~\ref{l:resetWrods}).

\begin{example}
Consider the automaton depicted in Figure~\ref{fig:aut}. This automaton is recurrent with the initial set of states $Q_0 = \set{q_x, q_a, q_b}$. 
For condition (1) from the definition of recurrent automata, observe that for every state $q$ we have $\widehat{\delta}(q, abab) = Q_0$. 
For condition (2), observe that $\widehat{\delta}(Q_0, b) = Q_0$, $\widehat{\delta}(Q_0, a) = \set{q_a, q_b}$ and $\widehat{\delta}(\set{q_a, q_b}, a) = \widehat{\delta}(\set{q_a, q_b}, b) =  Q_0$. 
The automaton would also be recurrent in the case of $Q_0=\set{q_a, q_b}$, but not in any other case.

Consider an automaton $\aut$ depicted below:

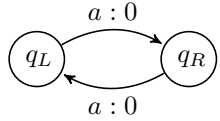
\begin{center}
{
\centering
\begin{tikzpicture}[->,>=stealth',shorten >=1pt,auto,node distance=2.0cm,semithick]
  \tikzstyle{every state}=[fill=white,draw=black,text=black,minimum size=0.4cm]
  \tikzset{
   datanode/.style = {
    draw,
    circle,
    text width=0.30cm,
    inner ysep = +0.4em},
    labelnode/.style = {
    draw, 
    rounded corners,
    align=center,
    fill=white}
  }

\foreach \lab\lX/\lY/\name/\kind/\cost/\time/\attr in 
         {A/0/0/$q_L$/square/0/10/5,
          B/2/0/$q_R$/walk/0/100/10
         }
{
  
  \node[datanode] (\lab) at (\lX,\lY)  
  {
   \name
  };
  
}

\draw  (A) edge[bend left] node[above] {$a:0$}  (B);
\draw  (B) edge[bend left] node {$a:0$} (A);

\end{tikzpicture}
}
 \end{center}

The automaton $\aut$ is recurrent if the set of initial states is either $\{q_L\}$ or $\{q_R\}$, but not in the case of $\{q_L, q_R\}$. 
Indeed, if we pick $q_L$ (resp., $q_R$) we can never reach the whole set $\{q_L, q_R\}$. 
This realizes our intuition that runs that start in $q_L$ and $q_R$ will never synchronize.
\end{example}

We discuss  properties of recurrent automata.
For every $\aut$ that is strongly connected as a graph there exists a set of initial states $T$ with which it becomes recurrent.
Indeed, consider $\aut$ as an unweighted $\omega$-automaton and construct a deterministic $\omega$-automaton $\aut^D$ through the power-set construction applied to $\aut$.
Observe that $\aut^D$ has a single bottom strongly-connected component (BSCC), i.e., a strongly connected component such that there are no transitions leaving that component. 
The set $Q_0$ belongs to that BSCC. 
Conversely, for any strongly connected automaton $\aut$, if $Q_0$ belongs to the BSCC of $\aut^D$, then $\aut$ is recurrent.

Observe that for a recurrent automaton $\aut$ the probability of words accepted by $\aut$  is either $0$ or $1$. 
Now, for a word $w$ consider a sequence of reachable sets of states $\Pi_w$ defined as $Q_0, \widehat{\delta}(Q_0, w[1]), \widehat{\delta}(Q_0, w[2]), \ldots$
Since $\aut^D$ has a single BSCC containing $Q_0$, all sets of $\Pi_w$ belong to that BSCC and hence 
either for almost all words $w$, the sequence $\Pi_w$ eventually contains only empty sets or
for all words $w$, the sequence $\Pi_w$ consists of non-empty sets only.
Observe that $\aut$ has an infinite run on $w$ if and only if $\Pi_w$ consists of non-empty sets.
It follows that the probability of the set of words having any infinite run in $\aut$ is either $0$ or $1$.
\medskip

While Markov chains generate words letter by letter, to define a run of a word of the minimal value we need to have the completely generated word, i.e., the optimal transition at some position $i$ may depend on some positions $j>i$ in the word.
This precludes application of standard techniques for probabilistic verification, which rely on the fact that the word and the run on it are generated simultaneously~\cite{DBLP:conf/focs/Vardi85,DBLP:journals/jacm/CourcoubetisY95,BaierBook}. 

\smallskip
\Paragraph{Key ideas} Our main idea is to change the non-determinism to \emph{bounded look-ahead}. 
This must be inaccurate, as the expected value of a deterministic automaton with bounded look-ahead is always rational, whereas Theorem~\ref{th:irrational} shows that the values of non-deterministic automata may be irrational. 
Nevertheless, we show that bounded look-ahead is sufficient to \emph{approximate} the probabilistic questions for recurrent automata (Lemma~\ref{l:convergence}). 
Furthermore, the approximation can be done effectively (Lemma~\ref{l:jumpingRuns}), which in turn
gives us an exponential-time approximation algorithm for recurrent automata (Lemma~\ref{l:singleSCC}). 
Then, we comment on the extension to all distributions given by Markov chains (Section~\ref{s:non-uniform}). 
Finally, we show the proof for all $\flimavg$-automata over probability measures given by Markov chains 
(Theorem~\ref{th:approximation-limavg}).

\subsection{Nearly-deterministic approximations}
 
\Paragraph{Jumping runs} Let $k>0$ and let $N_k$ be the set of natural numbers not divisible by $k$.
A \emph{$k$-jumping run} $\xi$ of $\aut$ on a word $w$ is an infinite sequence of states such that 
for every position $i \in N_k$ we have $(\pi[i-1],w[i],\pi[i]) \in \delta$.

An $i$-th \emph{block} of a $k$-jumping run is a sequence $\xi[ki, k(i+1)-1] $; within a block the sequence $\xi$ is consistent with transitions of $\aut$. The positions $k,2k, \ldots \notin N_k$ are \emph{jump} positions, where the sequence $\xi$ need not obey the transition relation of $\aut$.

The cost $\cost$ of a transition of a $k$-jumping run $\xi$ within a block is defined as usual, while the cost of a jump is defined as the minimal weight of $\aut$. 
The value of a $k$-jumping run $\xi$ is defined as the limit average computed for such costs.

\Paragraph{Optimal and block-deterministic jumping runs}
We say that a $k$-jumping run $\xi$  on a word $w$ is \emph{optimal} if its value is the infimum over values of all $k$-jumping runs on $w$.
We show that optimal $k$-jumping runs can be constructed nearly deterministically, i.e., only looking ahead to see the whole current block.

For every $S \subseteq Q$ and $u \in \Sigma^k$ we fix a run $\xi_{S,u}$ on $u$ starting in one of states of $S$, which has the minimal average weight.
Then, given a word $w \in \Sigma^{\omega}$, we define a $k$-jumping run $\xi$ as follows. 
We divide $w$ into $k$-letter blocks $u_1, u_2, \ldots$ and 
we put $\xi = \xi_{S_0, u_1} \xi_{S_1, u_2} \ldots$, where $S_0 = \set{q_0}$ and for $i>0$, $S_i$ is the set of states reachable from $q_0$ on the word $u_1 \ldots u_i$.
The run $\xi$ is a $k$-jumping run and it is indeed optimal. 
We call such runs \emph{block-deterministic} --- they can be constructed based on finite memory --- the set of reachable states $S_i$ and the current block of the input word.

Since all runs of $\aut$ are in particular $k$-jumping runs, the value of (any) optimal $k$-jumping run on $w$ is less or equal to $\aut(w)$. We show that for recurrent $\flimavg$-automata, the values of $k$-jumping runs on $w$ converge to $\aut(w)$ as $k$ tends to infinity. To achieve this, we construct a run of $\aut$ which tries to ``follow'' a given jumping run, i.e., 
after almost all jump positions it is able to synchronize with the jumping run quickly.

\Paragraph{Proof plan} Let $k>0$. Consider a word $w$ and some optimal $k$-jumping run $\optRun$ on $w$. 
We construct a run $\folRun$ of $\aut$ in the following way. Initially, both runs start in some initial state $q_0$ and coincide. 
However, at the first jump position $\optRun$ may take a move that is not a transition of $\aut$. 
The run $\folRun$ attempts to synchronize with $\optRun$, i.e., to be at the same position in the same state, and 
then repeat transitions of $\optRun$ until the end of the block. Then, in the next block, regardless of whether $\folRun$ managed to synchronize with $\optRun$ or not, we repeat the process. 
We say  that a run $\folRun$ constructed in such a way is a \emph{run following} $\optRun$. 

In the following Lemma~\ref{l:resetWrods}, we show that for $m \in \N$ large enough, with high probability, the run $\folRun$ synchronizes with $\optRun$ within $m$ steps.
We then show that if $m$ is large enough and $k$ is much larger than $m$, then the values of runs $\folRun$ and $\optRun$ differ by less than $\epsilon$ (Lemma~\ref{l:convergence}).

Let $q$ be a state of $\aut$ and $u$ be a finite word.
We say that a word $v$ \emph{saturates} the pair $(q, u)$, if the set of reachable states from $q$ over $v$ equals all the states reachable over $uv$ from the initial states, i.e., 
$\widehat{\delta}(Q_0, uv) = \widehat{\delta}(q, v)$. 

\begin{example}
Consider the automaton from Figure~\ref{fig:aut} with $Q_0=Q$. 
For any $(q, u)$, any word that contains the infix $abab$ saturates $(q, u)$, as
$\widehat{\delta}(Q_0, uv'abab) = \widehat{\delta}(q, v'abab)= Q_0$ for any $v'$.
\end{example}

Observe that in the above, the probability that a random word of a length $4\ell$
does not saturate $(q,u)$ is bounded by $(1-\frac{1}{16})^\ell$. So the probability that a random word $v$ saturates $(q, u)$ quickly tends to $1$ with $|v|$. The next lemma shows that this is not a coincidence.

\begin{lemma}
\label{l:resetWrods}
Let $\aut$ be an NFA, $u$ be a finite word, and $q \in \widehat{\delta}(Q_0, u)$.
For every $\Delta >0$ there exists a natural number $\ell=2^{2^{O(|\aut|)}} \log (\rev{\Delta})$ such that over the uniform distribution on $\Sigma^{\ell}$ we have
$\prob(\{v \in \Sigma^{\ell} \mid v\text{ saturates } (q,u) \}) \geq 1 - \Delta$.
\end{lemma}
\begin{proof}
First, observe that there exists a word $v$ saturating $(q,u)$.
Let $S = \widehat{\delta}(Q_0, u)$. Then, $q \in S$.
Since $\aut$ is recurrent, there exists a word $\alpha$ such that $Q_0 = \widehat{\delta}(q, \alpha)$. 
It follows that $S = \widehat{\delta}(q, \alpha u)$.
Since $q \in S$, we have $\widehat{\delta}(q, \alpha u) \subseteq \widehat{\delta}(S, \alpha u)$.
It follows that for $i\geq 0$ we have  
$\widehat{\delta}(S, (\alpha u)^i )  = \widehat{\delta}(q, (\alpha u)^{i+1}) \subseteq \widehat{\delta}(S, (\alpha u)^{i+1})$.
Therefore, for some $i >0$ we have $\widehat{\delta}(q, (\alpha u)^{i}) = \widehat{\delta}(S, (\alpha u)^{i})$, i.e., the word $(\alpha u)^{i}$ saturates $(q,u)$.

Now, we observe that there exists a saturating word that is exponentially bounded in $|\aut|$. 
We start with the word $v_0$ equal $(\alpha u)^{i}$ and we pick any two positions $k < l$ such that
$\widehat{\delta}(q, v_0[1,k]) = \widehat{\delta}(q, v_0[1,l])$ and 
$\widehat{\delta}(S, v_0[1,k]) = \widehat{\delta}(S, v_0[1,l])$. 
Observe that for $v_1$ obtained from $v_0$ by removal of $v[k+1, l]$,
the reachable sets do not change, i.e., $\widehat{\delta}(q, v_0) = \widehat{\delta}(q, v_1)$ and
$\widehat{\delta}(S, v_0) = \widehat{\delta}(S, v_1)$. We iterate this process until there are no such positions.
The resulting word $v'$ satisfies
$\widehat{\delta}(S, v') =  \widehat{\delta}(q, v')$. 
Finally, each position $k$ of $v'$ defines the unique pair $(\widehat{\delta}(q, v_0[1,k]), 
\widehat{\delta}(S, v_0[1,k]))$ of subsets of $Q$. Therefore, the length of $v'$ is bounded by $2^{2\cdot|Q|}$.

We have shown above that for every pair $(q,u)$ there exists a saturating word $v_{q,u}$ of length bounded by $N = 2^{2\cdot|Q|}$.
The probability of the word $v_{q,u}$ is $p_0 = 2^{-O(N)}$. 
Let $\ell = \frac{1}{p_0} \cdot \log (\rev{\Delta})$; we show that the probability that $(q, u)$ is not saturated by a word from $\Sigma^{N \cdot \ell}$
is at most $\Delta$. 
Consider a word $x \in \Sigma^{N \cdot \ell}$. We can write it as $x = x_1 \ldots x_{\ell}$, where all words $x_k$ have length $N$.
If $x_k$ saturates $(q,u x_1 \ldots x_{k-1})$, then $x_1 \ldots x_k$ (as well as $x$) saturates $(q,u)$. 
Therefore, the word $x$ does not saturate $(q_u)$ if for all $1 \leq k \leq \ell$, $x_k$ does not saturate $(q, u x_1 \ldots x_{k-1})$.
The probability that $x \in \Sigma^{N \cdot \ell}$ does not saturate $(q,u)$ is at most $(1 - p_0)^{\ell} \leq  (\frac{1}{2})^{\log (\rev{\Delta})} \leq \Delta$. 
\end{proof}

Finally, we show that for almost all words the value of an optimal $k$-jumping run approximates 
 the values of the word. 

\begin{lemma}
\label{l:convergence}
Let $\aut$ be a recurrent $\flimavg$-automaton.
For every $\epsilon \in \Q^+$, there exists $k$ such that
for almost all words $w$, the value $\aut(w)$ and the value of an optimal $k$-jumping run on $w$ differ by at most $\epsilon$.
The value $k$ is doubly-exponential in $|\aut|$ and polynomial in $\rev{\epsilon}$.
\end{lemma}
\begin{proof}
By Lemma~\ref{l:resetWrods}, for all $\Delta >0$,  $\ell = 2^{2^{O(|\aut|)}} \log (\rev{\Delta})$, and all $k > \ell$, 
the probability that the run $\folRun$ synchronizes with 
an optimal $k$-jumping run $\optRun$ within $\ell$ steps in a block is at least $1 - \Delta$.

Consider some $k>\ell$ and an optimal $k$-jumping run $\optRun$ that is block-deterministic.
Observe that the run $\folRun$ of $\aut$ following $\optRun$ is also block-deterministic.

Consider a single block $\optRun[i, i+k-1]$. By Lemma \ref{l:resetWrods}, 
the probability that $\folRun[i+\ell-1]=\optRun[i+\ell-1]$ is at least $1-\Delta$.
In such a case, the sum of costs on that block of $\folRun$ exceeds $\optRun$ by at most $D \cdot \ell$, 
where $D$ is the difference between the maximal and the minimal weight in $\aut$.
Otherwise, if $\folRun$ does not synchronize, we bound the difference of the sums of values on that block by 
the maximal possible difference $D \cdot k$. 

Since runs are block-deterministic, synchronization of $\folRun$ and $\optRun$ satisfies the Markov property; it 
depends only on the current block and the set of states $S$ reachable on the input word until the beginning of the current block.
We observe that as  $\aut$ is recurrent, the corresponding Markov chain, whose states are reachable sets of states  $S$ of $\aut$, has only a single BSCC.
Therefore, for almost all words, the average ratio of $k$-element blocks, in which
   $\folRun$ synchronizes with $\optRun$ within $\ell$ steps, is $1 - \Delta$.
We then conclude that for almost all words the difference between $\folRun$ and $\optRun$  is bounded by $\gamma = \frac{(1 - \Delta)\cdot (D \cdot \ell) + \Delta \cdot (D \cdot k)}{k}$. 
Observe that with $\Delta < \frac{\epsilon}{2\cdot D}$ and $k > \frac{2\cdot D \cdot \ell}{\epsilon}$, the value $\gamma$ is less than $\epsilon$.
\end{proof}

\subsection{Random variables}

Given a recurrent $\flimavg$-automaton $\aut$ and $k>0$, 
we define a function $\gKA : \Sigma^{\omega} \to \R$ such that
$\gKA(w)$ is the value of some optimal $k$-jumping run $\optRun$ on $w$.
We can pick $\optRun$ to be block-deterministic and 
hence $\gKA$ corresponds to a Markov chain $\exactMarkov$.
More precisely, we define $\exactMarkov$ labeled by $\Sigma^{k}$ such that 
for every word $w$,
the limit average of the path in $\exactMarkov$ labeled by blocks of $w$ (i.e., blocks $w[1,k] w[k+1, 2k] \ldots$) equals $\gKA(w)$.  
Moreover, the distribution of blocks $\Sigma^k$ is uniform and hence $\exactMarkov$ corresponds to $\gKA$ over the uniform distribution over $\Sigma$.
The Markov chain $\exactMarkov$ is a labeled weighted Markov chain~\cite{filar}, such that
its states are all subsets of $Q$, the set of states of $\aut$. 
For each state $S \subseteq Q$ and $u \in \Sigma^k$, the Markov chain $\markov$ has an edge $(S,\widehat{\delta}(S,u))$ 
of probability $\frac{1}{|\Sigma|^k}$.
The weight of an edge $(S,S')$ labeled by $u$ is the minimal average of weights of any run from some state of $S$ to some state of $S'$ over the word $w$.

We have the following:

\begin{lemma}
\label{l:recurrentMeasurable}
Let $\aut$ be a recurrent $\flimavg$-automaton and $k>0$.
(1)~The functions $\gKA$ and $\valueL{\aut}$  are random variables.
(2)~For almost all words $w$ we have $\gKA(w) = \expected(\gKA)$ and $\valueL{\aut}(w) = \expected(\valueL{\aut})$.
\end{lemma}
\begin{proof}
Since $\aut$ is recurrent, $\exactMarkov$ has a single BSCC and hence $\exactMarkov$ and $\gKA$ return the same value for almost all words~\cite{filar}. 
This implies that the preimage through $\gKA$ of each set has measure $0$ or $1$, and hence  $\gKA$  is measurable~\cite{feller}.
Lemma~\ref{l:convergence} implies that (measurable functions) $\gKA$ converge to $\valueL{\aut}$ with probability $1$, and hence $\valueL{\aut}$ is measurable~\cite{feller}.
As the limit of $\gKA$, $\valueL{\aut}$ also has the same value for almost all words.
\end{proof}

\begin{nremark}\label{r:ultimatelyperiodic}
The automaton $\aut$ from the proof of Theorem \ref{th:irrational} is recurrent (it resets after each $\$$), so the value of $\aut$ on almost all words is irrational. Yet, for every ultimately periodic word $vw^\omega$, the value of 
$\aut$ is rational. This means that while the expected value is realized by almost all words, it is not realized by any ultimately periodic word.
\end{nremark}

\subsection{Approximation algorithms}
We show that the expected value of $\gKA$ can be efficiently approximated.
The approximation is exponential in the size of $\aut$, but only logarithmic in $k$ (which is doubly-exponential due to Lemma~\ref{l:convergence}). 

\newcommand{\newh}[1]{\tilde{h}^{#1}}
\newcommand{\mfloor}[2]{\lfloor#1\rfloor_{#2}}
To approximate the expected value of $\gKA$ we need to compute the expected value of $\aut$ over $k$-letter blocks.
Such blocks are finite and hence we consider $\aut$ as a finite-word automaton with the average  value function $\favgFin$.
More precisely, for $S$ being a subset of states of $\aut$, we define $\aut_{S}^{\fin}$ as a $\favgFin$-automaton over finite words as $\aut$, which initial states set to $S$ and 
all states accepting. We can approximate the expected value of $\aut_{S}^{\fin}$ over words $\Sigma^{k}$ in logarithmic time in $k$. 

\begin{lemma}
\label{l:approxFinExpected}
Let $\aut$ be a recurrent $\flimavg$-automaton, let $S, S'$ be subsets of states of $\aut$, and let $i > 0$.
We can approximate the expected value $\expected( \{ \aut_{S}^{\fin}(w) \mid |w| = 2^k $ and 
$ \widehat{\delta}(S,w) = S'\})$ within a given $\epsilon \in \Q^+$
in exponential time in $|\aut|$ and polynomial time in $k$ and $\rev{\epsilon}$.
\end{lemma}

\begin{proof}
\newcommand{\roundup}[1]{\left[#1\right]_{\epsilon_0}}
Let $h(q, w, q')$ be the infimum average weight over runs from $q$ to $q'$ over $w$.
Consider $\epsilon_0 =\frac{\epsilon}{k+1}$. 
Let $H=\set{j\cdot \epsilon_0 \mid j\in \Z} \cap (-|\aut|, |\aut|)$ be a finite set and 
let $\roundup{x}$ stand for the greatest number from $H$ not exceeding $x$.

Consider $i \in \set{0, \dots, k}$ and let $N = 2^i$. We define a function $\newh{i}: Q\times \Sigma^{N} \times Q \to H$  as follows. First, we define $\newh{0}(q, w, q')=\roundup{h(q, w, q')}$. 
Then, inductively, we define 
\[\newh{i+1}(q,w_1w_2, q') = \roundup{\min_{q'' \in Q} \frac{\newh{i}(q, w_1, q'') + \newh{i}(q'', w_2, q')}{2}}\]

\noindent We show by induction on $i$ that for all $i$, $q$, $q'$, $N = 2^i$ and $w\in \Sigma^{N}$ we have $|h(q, w, q') -\newh{i}(q, w, q') |\leq (i+1) \epsilon_0$. 
First, we comment on the deteriorating precision. 
Notice that $|h(q, w, q') -\newh{i}(q, w, q') |\leq \epsilon_0$ may not hold in general. 
Let us illustrate this with a simple toy example.
Consider $\epsilon_0=1$, $x \in (0, 1)$ and $y \in (1, 2)$. Then $\frac{x+y}{2}\in (\frac{1}{2}, \frac{3}{2})$, thus $\roundup{\frac{x+y}{2}}\in \set{0, 1}$.
However, knowing only $\roundup{x}$ and $\roundup{y}$, we cannot asses whether the answer should be $0$ or $1$.
Therefore, when iterating the above-described procedure, we may lose some precision (up to one $\epsilon_0$ at each step); this is why we start with $\epsilon_0$ rather than $\epsilon$.

Now, we show by induction $|h(q, w, q') -\newh{i}(q, w, q') |\leq (i+1) \epsilon_0$. More precisely, we show that 
\begin{enumerate}[(1)]
\item $\newh{i}(q, w, q') \leq h(q, w, q') $ and 
\item $h(q, w, q') -\newh{i}(q, w, q') \leq (i+1) \epsilon_0$.
\end{enumerate}

The case $i=0$ follows from the definition of $\newh{0}(q, w, q')$.
Consider $i >0$ and assume that for all words $w$ of length $2^i$ the induction hypothesis holds. 
Consider $w = w_1 w_2$ and states $q, q'$.  There exists $q''$ such that 
$h(q, w, q') = h(q, w_1, q'') + h(q'', w_1, q')$. Then, due to induction assumption of (1) we have
$\newh{i-1}(q, w_1, q'')  \leq h(q, w_1, q'') $ and 
$\newh{i-1}(q'',w_2, q')  \leq h(q'', w_2, q') $. In consequence,  we get (1).

Now, to show (2), consider a state $s$ that realizes the minimum from the definition of $\newh{i}(q, w, q')$.
There are numbers $a,b \in \Z$ such that $\newh{i-1}(q, w_1, s) = a\epsilon_0$, 
and $\newh{i-1}(s, w_2, q') = b\epsilon_0$.

Then, $h(q,w,q') \leq \frac{h(q,w_1,s) + h(s,w_2,q)}{2}$ and we have
\[
h(q,w,q')  - \newh{i}(q, w, q') \leq 
\frac{h(q,w_1,s) + h(s,w_2,q)}{2} - \roundup{\frac{(a+b)\epsilon_0}{2}} 
\]
Observe that $\roundup{\frac{(a+b)\epsilon_0}{2}} =  \frac{(a+b)\epsilon_0}{2}$ if $a+b$ is even and
$
\roundup{\frac{(a+b)\epsilon_0}{2}} =  \frac{(a+b)\epsilon_0}{2} - \frac{\epsilon_0}{2}$ otherwise.
This gives us the following inequality
\[
h(q,w,q')  - \newh{i}(q, w, q') \leq \frac{(h(q,w_1,s)  -a\epsilon_0)+ (h(s,w_2,q)-b\epsilon_0)}{2} + \frac{\epsilon_0}{2} 
\]
Due to the induction hypothesis (2) we have $h(q, w_1, q'') - a\epsilon_0 \leq i \epsilon_0$
$h(q'', w_2, q') - b\epsilon_0 \leq i \epsilon_0$ and it gives us (2).

We cannot compute the functions $\newh{i}$ directly (in reasonable time), because there are too many words to be considered.
However, we can compute them symbolically. 
Define the \emph{clusterization function} $c^i$ as follows. Let $N = 2^i$. For each function $f \colon Q\times \Sigma^N \times Q \to H$ we define $c^i(f) = |\set{w \mid \forall q, q' .\newh{i}(q, w, q')=f(q, w, q')}|$. 
Basically, for each function $f$, clasterization counts the number of words realizing $f$ though functions $\newh{i}(\cdot, w, \cdot)$.

The function $c^0$ can be computed directly. Then, $c^{i+1}(f)$ can be computed as the sum of $c^{i}(f_1) \cdot c^{i}(f_2)$ over all the functions $f_1, f_2$ such that $f=f_1 * f_2$, where
$h_1 * h_2(q, q'')=\roundup{\min_{q' \in F_2} \frac{h_1(q,q') + h_2(q',q'')}{2}}$. 

It follows that we can compute the $k$-clusterization in time exponential in $|\aut|$, polynomial in $\rev{\epsilon}$ and $k$.
The desired expected valued can be derived from the $k$-clusterization in the straightforward way.
\end{proof}

In consequence, we can approximate the expected value of $\gKA$ in exponential time in $|\aut|$ but logarithmic in $k$, which is important as $k$ may be doubly-exponential in $|\aut|$ (Lemma~\ref{l:convergence}).

\begin{lemma}
\label{l:jumpingRuns}
Given a recurrent $\flimavg$-automaton $\aut$,  $k=2^l$ and $\epsilon \in \Q^+$, the expected value $\expected(\gKA)$ can be approximated up to $\epsilon$
in exponential time in $|\aut|$, logarithmic time in $k$
 and polynomial time in $\rev{\epsilon}$. 
\end{lemma}
\begin{proof}
Recall that the expected values of $\exactMarkov$ and $\gKA$ coincide. 
Observe that $\exactMarkov$ can be turned into a weighted Markov chain $\compMarkov$ over the same set of states with one edge between any pair of states as follows. 
For an edge $(S,S')$, we define its probability as 
 $\frac{1}{|\Sigma|^k}$ multiplied by the number of edges from $S$ to $S'$ with positive probability in $\exactMarkov$ and the weight of  
 $(S,S')$ in $\compMarkov$ is the average of the weights of all such the edges in $\exactMarkov$, i.e., 
 the weight of $(S,S')$ is $\expected( \{ \aut_{S}^{\fin}(w) \mid w \in \Sigma^k $ and $ \widehat{\delta}(S,w) = S'\})$ (see Lemma~\ref{l:approxFinExpected}).
Observe that the expected values of $\exactMarkov$ and  $\compMarkov$ coincide. 

Having the Markov chain $\compMarkov$, we can compute its expected value in polynomial time~\cite{filar}.
Since $\compMarkov$ has the exponential size in $|\aut|$, we can compute it in exponential time in $|\aut|$.
However, we need to show how to construct $\compMarkov$. In particular, computing
$\expected( \{ \aut_{S}^{\fin}(w) \mid w \in \Sigma^k $ and $ \widehat{\delta}(S,w) = S'\})$ can be computationally expensive as $k$ can be doubly-exponential in $|\aut|$ (Lemma~\ref{l:convergence}). 
Still, due to Lemma~\ref{l:approxFinExpected}, we can approximate $\expected( \{ \aut_{S}^{\fin}(w) \mid w \in \Sigma^k $ and $ \widehat{\delta}(S,w) = S'\})$ in exponential time in $|\aut|$, logarithmic time in $k$
 and  polynomial  time in $\epsilon$.
Therefore, we can compute a Markov chain $\apxMarkov$ with the same structure as $\compMarkov$ and such that 
for every edge  $(S,S')$ the weight of $(S,S')$ in $\apxMarkov$ differs from the weight in $\compMarkov$ by at most $\epsilon$.
Therefore, the expected values of  $\apxMarkov$ and $\compMarkov$ differ by at most ${\epsilon}$.
\end{proof}

Lemma~\ref{l:convergence} and Lemma~\ref{l:jumpingRuns} give us approximation algorithms for the expected value and the distribution of recurrent automata over the uniform distribution: 
\begin{lemma}
\label{l:singleSCC-uniform}
Given a recurrent $\flimavg$-automaton $\aut$, $\epsilon \in \Q^+$ and $\lambda \in \Q$, 
we can compute $\epsilon$-approximations
of the distribution $\distrib_{\aut}(\lambda)$ and  the expected value $\expected(\aut)$
with respect to the uniform measure 
in exponential time in $|\aut|$ and polynomial time in $\rev{\epsilon}$.
\end{lemma}

\begin{proof}
For uniform distributions, by Lemma~\ref{l:convergence}, for every $\epsilon > 0$, there exists 
$k$ such that $|\expected(\aut) - \expected(\gKA) | \leq \frac{\epsilon}{2}$.
The value $k$ is doubly-exponential in $|\aut|$ and polynomial in $\rev{\epsilon}$.
Then, by 
Lemma~\ref{l:jumpingRuns}, we can compute $\gamma$ such that $|\gamma - \expected(\gKA)| \leq  \frac{\epsilon}{2}$ in exponential time in $|\aut|$ and polynomial in $\rev{\epsilon}$.
Thus, $\gamma$ differs from $\expected(\aut)$ by at most $\epsilon$.
Since almost all words have the same value, we can approximate $\distrib_{\aut}(\lambda)$ by comparing $\lambda$ with $\gamma$, i.e.,
$1$ is an $\epsilon$-approximation of $\distrib_{\aut}(\lambda)$ if $\lambda \leq \gamma$, and otherwise $0$ is an $\epsilon$-approximation of $\distrib_{\aut}(\lambda)$.
\end{proof}

\subsection{Non-uniform measures}
\label{s:non-uniform}

We briefly discuss how to adapt Lemma~\ref{l:singleSCC-uniform} to all measures given by Markov chains. We sketch the main ideas.

\noindent \emph{Key ideas}. 
Assuming that (a variant of) Lemma~\ref{l:resetWrods} holds for any probability measure given by a Markov chain, the proofs of 
Lemmas~\ref{l:convergence}, \ref{l:jumpingRuns} and \ref{l:singleSCC} can be easily adapted.
Therefore we focus on adjusting Lemma~\ref{l:resetWrods}.

Observe that if a Markov chain $\markov$ produces all finite prefixes $u \in \Sigma^*$ with non-zero probability, then the proof of  Lemma~\ref{l:resetWrods} can be straightforwardly adapted.
Otherwise, if some finite words cannot be produced by a Markov chain $\markov$, then Lemma~\ref{l:resetWrods} may be false.
However, if there are words $w$ such that  $\aut$ has an infinite run on $w$, but  $\markov$ does not emit $w$, we can restrict $\aut$ to reject such words.
Therefore, we assume that for every word $w$, if $\aut$ has an infinite run on $w$, then $\markov$ has an infinite path with non-zero probability on $w$ ($\markov$ emits $w$).
Then, the current proof of Lemma~\ref{l:resetWrods} can be straightforwardly adapted to the probability measure given by $\markov$. 
In consequence, we can compute $\distrib_{\markov, \aut}(\lambda)$ and $\expected_{\markov}(\aut)$
in exponential time in $|\aut|$ and polynomial time in  $|\markov|$ and $\rev{\epsilon}$.

More precisely, we first observe that we may enforce $\markov$ to be ``deterministic'', i.e., for all states $s$ and letters $a$ at most one outgoing transition labeled with $a$ has positive probability.
We can determinise $\markov$ by extending the alphabet $\Sigma$ to $\Sigma \times S$, where $S$ is the set of states of $\markov$. 
The second component in  $\Sigma \times S$ encodes the target state in the transition. 
Observe that $\aut$ can be extended to the corresponding automaton $\aut'$ over $\Sigma \times S$ by cloning transitions, i.e., for every transition $(q,a,q')$,  the automaton $\aut'$
has transitions  $(q,(a,s),q')$ for every $s \in S$ (i.e., $\aut'$ ignores the state of $\markov$). For such a deterministic Markov chain $\markov'$, we define a deterministic $\omega$-automaton $\aut_{\markov}$ that accepts words emitted  
by $\markov'$. Finally, we consider the automaton $\aut^R = \aut_{\markov} \times \aut'$, which has infinite runs only on words that are emitted by $\markov'$.
Therefore, as we discussed, we can adapt the proof of Lemma~\ref{l:singleSCC-uniform} in such a case and  compute $\distrib_{\markov', \aut^R}(\lambda)$ and
$\expected_{\markov'}(\aut^R)$ (in exponential time in $|\aut^R|$, polynomial time in  $|\markov|$ and $\rev{\epsilon}$; notice that $|\aut^R|$ is polynomial in $|\aut|$).
Finally, observe that
 $\distrib_{\markov, \aut}(\lambda) =  \distrib_{\markov', \aut^R}(\lambda)$ and
 $\expected_{\markov}(\aut) = \expected_{\markov'}(\aut^R)$.
In consequence, we have the following:

\begin{lemma}
\label{l:singleSCC}
Given a recurrent $\flimavg$-automaton $\aut$, Markov chain $\markov$, $\epsilon  \in \Q^+$ and $\lambda \in \Q$, we can compute $\epsilon$-approximations
of the distribution $\distrib_{\markov, \aut}(\lambda)$ and  the expected value $\expected_{\markov}(\aut)$
in exponential time in $|\aut|$ and polynomial time in $|\markov|$ and $\rev{\epsilon}$.
\end{lemma}

\section{Non-recurrent automata}
\label{s:non-recurrent}

We present the approximation algorithms for all non-deterministic $\flimavg$-automata over measures given by Markov chains. 

\begin{theorem}
\label{th:approximation-limavg}
(1)~For a non-deterministic $\flimavg$-automaton $\aut$ the function $\valueL{\aut} : \Sigma^{\omega} \to \R$ is measurable.
(2)~Given a non-deterministic $\flimavg$-automaton $\aut$, Markov chain $\markov$, $\epsilon \in \Q^+$, and $\lambda \in \Q$, 
 we can $\epsilon$-approximate the distribution $\distrib_{\markov, \aut}(\lambda)$ and the expected value $\expected(\aut)$
in exponential time in $|\aut|$ and polynomial time in $|\markov|$ and $\rev{\epsilon}$.
\end{theorem}

\begin{proof}
\newcommand{\Mpath}{\rho}
\newcommand{\autDet}{\aut^D}
Consider $\aut$ as an $\omega$-automaton. 
It has no acceptance conditions and hence we can determinise it with the standard power-set construction to a deterministic automaton $\autDet$.
Then, we construct a Markov chain $\markov \times \autDet$, compute all its BSCCs $R_1, \ldots, R_k$ along with the probabilities $p_1, \ldots, p_k$ of reaching each of these sets.
This can be done in polynomial time in $\markov \times \autDet$~\cite{filar,BaierBook}, and hence polynomial in $\markov$ and exponential in $\aut$.
Let $H_1, \ldots, H_k$ be sets of paths in $\markov \times \autDet$ such that for each $i$, all $\Mpath \in H_i$ eventually reach $R_i$ and stay there forever. 
Observe that each $H_i$ is a Borel set; the set $H_i^p$ of paths that stay in $R_i$ past position $p$ is closed and $H_i = \bigcup_{p\geq 0} H_i^p$.
It follows that each $H_i$ is measurable. 
We show how to compute an $\epsilon$-approximation of the conditional expected value $\expected_{\markov}(\aut \mid H_i)$.

Consider a BSCC $R_i$. The projection of $R_i$ on the first component $R_i^1$ is a BSCC in $\markov$ and the projection on the second component $R_i^2$ is an SCC of $\aut_D$.
Let $(s,A) \in R_i$. If we fix $s$ as the initial state of $R_i^1$ and $A$ as the initial state of $R_i^2$, then $R_i$ are all reachable states of $R_i^1 \times R_i^2$.  
The set $R_i^2$ consists of the states of $\autDet$, which are subsets of states of $\aut$. 
Therefore, the union $\bigcup R_i^2$ is a subset of states of $\aut$ and it consists of some SCCs $S_1,  \ldots, S_m$ of $\aut$.
All these SCCs are reachable, but it does not imply that there is a run of $\aut$ that stays in $S_j$ forever. We illustrate that in the following example.

Consider the automaton $\aut$ presented in Figure~\ref{fig:autTwo}, where $q_I$ is the initial state, and a single-state Markov chain $\markov$ generating uniform distribution.
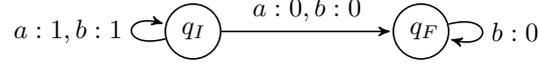
\begin{figure}
\centering
\begin{tikzpicture}[->,>=stealth',shorten >=1pt,auto,node distance=2.0cm,semithick]
  \tikzstyle{every state}=[fill=white,draw=black,text=black,minimum size=0.4cm]
  \tikzset{
   datanode/.style = {
    draw,
    circle,
    text width=0.3cm,
    inner ysep = +0.4em},
    labelnode/.style = {
    draw, 
    rounded corners,
    align=center,
    fill=white}
  }

\foreach \lab\lX/\lY/\name/\kind/\cost/\time/\attr in 
         {B/0/0/$q_I$/square/0/10/5,
          A/3/0/$q_F$/bus/10/15/-2}
{
  
  \node[datanode] (\lab) at (\lX,\lY)  
  {
   \name
  };
  
}

\draw  (B) edge node {$a:0, b:0$} (A);
\draw  (A) edge[loop right] node {$b:0$} (A);
\draw  (B) edge[loop left] node {$a:1, b:1$} (B);
\end{tikzpicture}
\caption{An automaton with a reachable SCC $q_F$ such that almost no runs stay in $q_F$ forever}
\label{fig:autTwo}
\end{figure}
Then, all paths in $\markov \times \autDet$ are eventually contained in $\markov \times \{Q\}$, i.e., the second component consists of all states of $\aut$. 
Still, if a word $w$ has infinitely many letters $a$, then $\aut$ has no (infinite) run on $w$ that visits the state $q_F$.
The set of infinite words that contain finitely many letters $a$ is countable and hence has probability $0$. 
Therefore, almost all words (i.e., all except for some set of probability $0$) have no run that visits the state $q_F$.

To avoid such pathologies, we divide SCCs into two types: \emph{permanent} and \emph{transitory}. 
More precisely, for a path $\Mpath$ in $\markov \times \autDet$ let $w_{\Mpath}$ be the word labeling $\Mpath$. 
We show that for each SCC $S_j$, one of the following holds:
\begin{itemize}
\item $S_j$ is \emph{permanent}, i.e., for almost all paths $\Mpath \in H_i$ (i.e., the set of paths of probability $1$), the automaton $\aut$ has a run on the word $w_{\Mpath}$ that eventually stays in $S_j$ forever, or
\item $S_j$ is \emph{transitory}, i.e., for almost all paths $\Mpath \in H_i$, the automaton $\aut$ has no run on $w_{\Mpath}$ that eventually stays in $S_j$.
\end{itemize}

Consider an SCC $S_j$. If $S_j$ is permanent, then it is not transitory. We show that if $S_j$  is not permanent, then it is transitory.
Suppose that $S_j$ is not permanent and consider any $(s,A) \in H_i$. 
Almost all paths in $H_i$ visit $(s,A)$ and since $S_j$ in not permanent,   
there exists an infinite path $\Mpath$ that visits $(s,A)$ and $\aut$ has no run on $w_{\Mpath}$ that stays in $S_j$ forever. 
Let $u$ be the suffix of $w_{\Mpath}$ that labels $\Mpath$ past some occurrence of $(s,A)$.
We observe that $\widehat{\delta}(A \cap S_j, u) = \emptyset$ and hence for some finite prefix $u'$ of $u$ we have $\widehat{\delta}(A \cap S_j, u') = \emptyset$.
Let $p$ be the probability that $\markov \times \autDet$ in the state $(s,A)$ generates a path labeled by $u'$. 
The probability that a path that visits $(s,A)$ at least $\ell$ times does not contain $(s,A)$ followed by labels $u'$ is at most $(1-p)^{\ell}$.
Observe that
for almost all paths in $H_i$, the state $(s,A)$ is visited infinitely often and hence almost all paths contain $(s,A)$ followed by labels $u'$ upon which the path leaves $S_j$.
Therefore, $S_j$ is transitory.

To check whether $S_j$ is permanent or transitory, observe that for any $(s,A) \in H_i$,
in the Markov chain $\markov \times \autDet$, we can reach the set $\markov \times \{\emptyset \}$ from $(s, A \cap S_j)$ if and only if
$S_j$ is transitory. The former condition can be checked in polynomial space.

We  mark each SCC $S_1, \ldots, S_k$ as permanent or transitory and
for every permanent SCC $S_j$, we compute 
an $\epsilon$-approximation of $\expected_{\markov}(\aut[S_j] \mid H_i)$, which is the expected value of $\aut$ under condition $H_i$ with the restriction to runs that
eventually stay in $S_j$.
Observe that an $\epsilon$-approximation of $\expected_{\markov}(\aut[S_j] \mid H_i)$ can be computed  
using Lemma~\ref{l:singleSCC}.
Indeed, we pick $(s, A) \in H_i$ and observe that $\aut$ restricted to states $S_j$ is recurrent (with an appropriate initial states). 
Finally, we pick the minimum $\gamma$ over the computed expected values $\expected_{\markov}(\aut[S_j] \mid H_i)$ and observe that almost all words in $H_i$ have value $\gamma$. 
It follows that $\expected_{\markov}(\aut \mid H_i) = \gamma$.

In each BSCC $R_i$, almost all words have value $\expected_{\markov}(\aut \mid H_i)$. 
As we discussed earlier, each $H_i$ is measurable, and hence the function $\valueL{\aut} : \Sigma^{\omega} \to \R$ is measurable.
Moreover, to approximate the distribution $\distrib_{\markov, \aut}(\lambda)$, 
we sum probabilities of $p_i$ of reaching the BSCCs $R_i$ over $R_i$'s such that the $\epsilon$-approximation of $\expected_{\markov}(\aut \mid H_i)$ is less or equal to $\lambda$.
Finally, we compute an $\epsilon$-approximation of $\expected_{\markov}(\aut)$ from $\epsilon$-approximations of conditional expected values $\expected_{\markov}(\aut \mid H_i)$ using the
identity $\expected_{\markov}(\aut) = \sum_{i=1}^k p_i \cdot \expected_{\markov}(\aut \mid H_i)$.
\end{proof}

\section{Determinising and approximating $\flimavg$-automata}
\newcommand{\autB}{\mathcal{B}}
For technical simplicity, we assume that the distribution of words is uniform. 
However, the results presented here extend to all distributions given by Markov chains. 

Recall that for the $\flimavg$ automata, the value of almost all words (i.e., all except for some set of words 
of probability $0$) whose optimal runs end up in the same SSC, is the same. 
This means that there is a finite set of values (not greater than the number of SSCs of the automaton) such that 
almost all words have their values in this set.

$\flimavg$-automata are not determinisable~\cite{quantitativelanguages}. 
We say that a non-deterministic $\flimavg$-automaton $\aut$ is \emph{weakly determinisable} if there 
is a deterministic $\flimavg$-automaton $\autB$ such that $\aut$ and $\autB$ have the same value over almost all words. 
From \cite{lics16} we know that deterministic automata return rational values for almost all words, so not all $\flimavg$-automata are weakly determinisable. 
However, we can show the following.

\begin{theorem}\label{t:determinisation}
A $\flimavg$-automaton $\aut$ is weakly determinisable if and only if 
it returns rational values for almost all words.
\end{theorem}

\begin{proof}[Proof sketch]
Assume an automaton $\aut$ with SSCs  $C_1, \dots, C_m$. 
For each $i$ let $v_i$ be defined as 
the expected value of $\aut$ when its set of initial states is $C_i$ and the run is bounded to stay in $C_i$.
If $\aut$ has no such runs for some $C_i$, then $v_i = \infty$.

We now construct a deterministic automaton $B$ with rational weights using the standard power-set construction. 
We define the cost function such that the cost of any transition from a state $Y$ is the minimal value $v_i$ such that $v_i$ is rational and $Y$ contains a state from $C_i$.
If there are no such $v_i$, then we set the cost to the maximal cost of $\aut$.
 Roughly speaking, $B$ tracks in which SSCs $A$ can be and the weight corresponds to the SSC with the lowest value.

To see that $B$ weakly determinises $A$ observe that for almost all words $w$, a run with the lowest value over $w$ ends in some SSC and its value then equals the expected value of this component,
which is rational as the value of this word is rational. 
\end{proof}  

A straightforward corollary is that every non-deterministic $\flimavg$-automaton can be weakly determinised by an $\flimavg$-automaton with real weights.

Theorem \ref{t:determinisation} does not provide an implementable algorithm for weak-determinisation, because of the hardness of computing the values $v_i$. 
It is possible, however, to approximate this automaton. 
We say that a deterministic $\flimavg$-automaton $B$ \emph{$\epsilon$-approximates} $\aut$ if for almost every word $w$ we have that $\valueL{B}(w)\in[\valueL{\aut}(w) - \epsilon, \valueL{\aut}(w) + \epsilon]$.

\begin{theorem}
\label{th:approximateDeterminisation}
For every $\epsilon>0$ and a non-deterministic $\flimavg$-automaton $\aut$, one can compute in exponential time a deterministic $\flimavg$-automaton that $\epsilon$-approximates $\aut$.
\end{theorem}

The proof of this theorem is similar to the proof of Theorem~\ref{t:determinisation}, except now it is enough to approximate the values $v_i$, which can be done in exponential time.

\Paragraph{Acknowledgements}
Our special thanks go to G\"unter Rote who pointed out an error in an earlier version of our running example.

\doclicenseThis

\bibliography{papers}
\end{document}